\documentclass{article}
\usepackage[utf8]{inputenc}
\usepackage[T1]{fontenc}
\usepackage{amsmath, amssymb}
\usepackage{hyperref}

\providecommand{\keywords}[1]
{
  \small	
  \textbf{\textit{Keywords---}} #1
}

\title{A comprehensive survey of Schwarzschild's original papers: Schwarzschild's trick and Einstein's s(h)tick}
\author{Galina Weinstein
  \thanks{Reichman University, The Efi Arazi School of Computer Science and The Department of Philosophy, University of Haifa.} 
  }

\begin{document}

\maketitle

\begin{abstract}
This paper examines Schwarzschild's key contributions to general relativity through his two papers. It focuses on his method for developing exterior and interior solutions. The study emphasizes Schwarzschild's ingenious methods and the implications of his solutions. The paper delves into the exchange of letters between Schwarzschild and Einstein, highlighting the collaborative nature of their scientific interaction.
Interestingly, despite presenting Schwarzschild's exact solutions to the Prussian Academy, Einstein exhibited a preference for his approximate methods in his 1916 review paper "The Foundation of the General Theory of Relativity." Contrary to common belief, the paper reveals that in 1916, Einstein's preference for approximate solutions over Schwarzschild's exact exterior solution was not due to a singularity concern. This finding dissociates Einstein's 1916 methodology from his later preoccupation with singularities, which became prominent only during his subsequent focus on unified field theory.
Thus, the research posits that factors other than singularities influenced Einstein's decision to stick with approximate methods in 1916. 
\end{abstract}

\keywords{Schwarzschild, Einstein, general relativity, unimodular coordinates, Mercury perihelion}

\tableofcontents

\section{Introduction}

This paper highlights the interplay between Karl Schwarzschild's mathematical ingenuity and Albert Einstein's physical approach in the foundational years of general relativity. Schwarzschild's contributions provided a means to overcome some limitations of Einstein's initial framework, while Einstein's broader perspective and methodology guided the theory's development.

Einstein's general theory of relativity is fundamentally oriented towards heuristic physical principles and thought experiments. His initial formulations and approximations from 1915-1916 reflected a more physical than mathematical orientation. 

Schwarzschild developed a clever method to address two major challenges in Einstein's theory of general relativity. His approach allowed for solving Einstein's 1915 field equations, which were restricted to unimodular coordinates. Additionally, it adhered to the four conditions Einstein had set forth in his pivotal 1915 paper on Mercury's perihelion advance, "Explanation of the Perihelion
Motion of Mercury from the General Theory of Relativity" \cite{Einstein15}. These conditions included static, spherically symmetric fields, satisfying specific metric conditions, and reverting to a flat metric at great distances.

Despite acknowledging Schwarzschild's exact solutions and presenting them to the Prussian Academy, Einstein persisted in using and refining his approximate methods in general relativity. When addressing Mercury's perihelion advance, Einstein employed an approximate solution within the framework of unimodular coordinates. His approximate solution included simplifying problems like Mercury's orbit and the bending of light rays in the Sun's gravitational field. 
This preference was particularly evident in his 1916 review paper, "The Foundation of the General Theory of Relativity" \cite{Einstein5}. 

Recognizing the restrictive nature of unimodular coordinates, Schwarzschild proposed finding an alternative condition that could simplify the field equations of general relativity \cite{CPAE8}, Doc 188.
While Einstein acknowledged the need to extend his theory beyond unimodular coordinates and eventually did so, his initial and primary development of the theory (1912-1915) heavily relied on these coordinates. This reliance was partly due to his preference for a physical rather than a purely mathematical methodology, which led him through various complexities and challenges in formulating his theory. 
Additionally, Einstein focused on ensuring that his theory converged with the Newtonian limit and adhered to the principle of energy-momentum conservation. Using an approximation methodology within unimodular coordinates was particularly significant in situations where gravitational fields were not strong, and the differences from Newtonian limits were minor. 

In this paper, I show Schwarzschild's pivotal role in enhancing and enriching the mathematical foundation of Einstein's work. Schwarzschild's contributions were instrumental in delving into the less intuitive yet highly mathematically rigorous aspects of Einstein's theory of general relativity. The collaboration and intellectual exchange between Einstein and Schwarzschild during 1915-1916 proved critical in advancing the theory. Their interaction enabled the theory of general relativity to evolve and attain its comprehensive capacity for describing intricate gravitational phenomena, thereby profoundly influencing Einstein's subsequent work.

\section{Schwarzschild and Einstein exchange letters} \label{1}

In his November 18, 1915, presentation on the perihelion of Mercury, Einstein derived approximate solutions to his vacuum field equations, starting with a basic approximation of a flat Minkowski space-time and then developing a first-order approximate solution \cite{Einstein15}:

\begin{equation} \label{Eq 29}
g_{\rho\sigma} = -\delta_{\rho\sigma} + \alpha\left(\frac{\partial^2 r}{\partial x_\rho \partial x_\sigma} - \frac{\delta_{\rho\sigma}}{r}\right) = -\delta_{\rho\sigma} - \alpha\frac{x_{\rho} x_{\sigma}}{r^3}, \quad g_{44} = 1 - \frac{\alpha}{r}. 
\end{equation}

The first term on the right-hand side, $-\delta_{\rho\sigma}$, represents the Minkowski metric component, indicating a flat spacetime background without gravitational effects. 
The diagonal matrix $(-1, 1, 1, 1)$ in Cartesian coordinates represents the Minkowski metric. 

The second term on the right-hand side, $\alpha\frac{x_{\rho} x_{\sigma}}{r^3}$, contributes to the metric tensor components that involve a constant $\alpha$ and the product of two coordinates $x_\rho$ and $x_\sigma$, divided by the cube of the radial distance from the Sun. 

The third term on the right-hand side, the $g_{44}$ component, can be written as:

\begin{equation}
g_{44} = 1 - \frac{2Gm}{r},    
\end{equation}

\noindent where the coefficient $\alpha$ is given by equation \eqref{Eq 73}, and is related to the strength of the gravitational field and the mass of the Sun (see section \ref{4}). 

In the weak-field limit, where the gravitational field is not too strong, the term $\frac{2Gm}{r}$ is small compared to $1$. This is consistent with the weak-field approximation where the gravitational potential $\Phi$ is introduced, and $g_{44}$ takes the form \cite{Einstein5}:

\begin{equation}
g_{44} = 1 + 2\Phi.    
\end{equation}

\noindent where $\Phi = -\frac{Gm}{r}$ in the case of a point mass.

The second term on the left-hand side, $\alpha\left(\frac{\partial^2 r}{\partial x_\rho \partial x_\sigma}\right)$, represents a small perturbation due to a weak gravitational field. $r$ denotes a radial coordinate, particularly in spherically symmetric spacetimes. $\frac{\partial^2 r}{\partial x_\rho \partial x_\sigma}$ is a second-order spatial derivative of the radial coordinate. 
In his 1915 papers, Einstein derived terms with second derivatives in the weak field regime \cite{Einstein14}. Indeed, second derivatives of the metric components naturally appear in the weak field approximation. 

Finally, the third term on the left-hand side, the term $\frac{\delta_{\rho\sigma}}{r}$: in the diagonal components of the metric tensor, this term would introduce a scaling factor that varies inversely with $r$. In the off-diagonal components (where $\rho \neq \sigma$ ), this term would contribute nothing since 
$\delta_{\rho\sigma}$ would be zero.

Einstein integrates spherical coordinates in his approximate solution \eqref{Eq 29}. Equation \eqref{Eq 29} mixes elements characteristic of both spherical and Cartesian coordinates. This can be seen in the way the spatial components and the time component of the metric tensor \( g_{\mu\nu} \) are expressed:
The term \( -(\delta_{\rho\sigma} + \alpha\frac{ x_{\rho} x_{\sigma}}{r^3}) \) for \( \mu = \rho, \nu = \sigma \) and \( \rho, \sigma = 1, 2, 3 \) uses \( x_{\rho} \) and \( x_{\sigma} \), which are typically Cartesian coordinates (or components thereof).
This part of the metric suggests mainly a Cartesian-like treatment of the spatial coordinates, especially with the use of \( \delta_{\rho\sigma} \) (the Kronecker delta) and the Cartesian product of coordinates \( x_{\rho} x_{\sigma} \).

The \( g_{44} = 1 - \frac{\alpha}{r} \) component, however, is more characteristic of a spherical coordinate system, particularly due to the presence of the radial distance \( r \).
This metric part is aligned with what we expect in a spherically symmetric spacetime.
Hence, the left-hand and the right-hand sides of equation \eqref{Eq 29} combine elements of both coordinate systems. 

Einstein transformed these terms in equation \eqref{Eq 29}: 

\begin{equation} \label{Eq 560}
-\delta_{\rho\sigma} + \alpha\left(\frac{\partial^2 r}{\partial x_\rho \partial x_\sigma} - \frac{\delta_{\rho\sigma}}{r}\right).    
\end{equation}

\noindent into these terms:

\begin{equation} \label{Eq 561}
-\delta_{\rho\sigma} - \alpha\frac{x_{\rho} x_{\sigma}}{r^3}, \quad g_{44} = 1 - \frac{\alpha}{r},    
\end{equation}

\noindent where Einstein defined: 

\begin{equation} \label{Eq 564}
r = \sqrt{x^2 + y^2 +z^2}.    
\end{equation}

Let us show how the transformation is performed from equation \eqref{Eq 560} to equation \eqref{Eq 561}. 
According to Einstein's four conditions (see section \ref{2}), we start by finding $\frac{\partial r}{\partial x}$:

\begin{equation}
\frac{\partial r}{\partial x} = \frac{\partial}{\partial x} \sqrt{x^2 + y^2 + z^2}.
\end{equation}
\vspace{1mm} %1mm vertical space 

\noindent Using the chain rule, this becomes:\footnote{Given the function \eqref{Eq 564}, using the chain rule:

\begin{equation} \label{Eq 365}
\frac{\partial r}{\partial x} = \frac{\partial r}{\partial u} \cdot \frac{\partial u}{\partial x}, \text{ where: }
\end{equation}

\begin{equation} \label{Eq 358}
u = x^2 + y^2 + z^2.    
\end{equation}

First, we find the first term on the right-hand side of equation \eqref{Eq 365}: 

\begin{equation} \label{Eq 356}
\frac{\partial r}{\partial u} = \frac{1}{2\sqrt{u}} \text{ since } r = \sqrt{u}.
\end{equation}

Next, we find the second term on the right-hand side of equation \eqref{Eq 365}:

\begin{equation} \label{Eq 359}
\frac{\partial u}{\partial x} = 2x,
\end{equation}

\noindent since $u$ is given by equation \eqref{Eq 358} and $y, z$ are constants with respect to $x$.

\noindent Now we combine equations \eqref{Eq 356} and \eqref{Eq 359} and plug them into equation \eqref{Eq 365}:

\begin{equation} \label{Eq 357}
\frac{\partial r}{\partial x} = \frac{1}{2\sqrt{u}} \cdot 2x = \frac{x}{\sqrt{u}}.
\end{equation}

\noindent We substitute equations \eqref{Eq 358} and  \eqref{Eq 564} back: 

\begin{equation}
\frac{\partial r}{\partial x} = \frac{x}{\sqrt{x^2 + y^2 + z^2}}. \quad \text{ And finally simplify:}
\end{equation}

\begin{equation}
\frac{\partial r}{\partial x} = \frac{x}{r}.
\end{equation}}

\begin{equation} \label{Eq 563}
\frac{\partial r}{\partial x} = \frac{1}{2} \cdot \frac{1}{\sqrt{x^2 + y^2 + z^2}} \cdot 2x = \frac{x}{\sqrt{x^2 + y^2 + z^2}} = \frac{x}{r}.
\end{equation}

\noindent Then, applying the product rule and the power rule and plugging equation \eqref{Eq 563} into the second derivative of $r$ with respect to $x$, we calculate the second derivative:

\begin{align}
\frac{\partial^2 r}{\partial x^2} &= \frac{\partial}{\partial x} \left( \frac{x}{r} \right) \\
&= \frac{\partial}{\partial x} \left( x \cdot \frac{1}{r} \right) \\
&= \frac{\partial x}{\partial x} \cdot \frac{1}{r} + x \cdot \frac{\partial}{\partial x}\left( \frac{1}{r} \right) \\
&= \frac{1}{r} + x \cdot \left( -\frac{x}{r^3} \right).
\end{align}

\noindent We have calculated that:

\begin{equation} \label{Eq 562}
\frac{\partial^2 r}{\partial x^2} = \frac{1}{r} + x \cdot \left( -\frac{x}{r^3} \right).
\end{equation}

\noindent Inserting equation \eqref{Eq 562} into \eqref{Eq 560} becomes:
\begin{equation}
-\delta_{\rho\sigma} + \alpha\left[\left(\frac{1}{r} + x \cdot \left( -\frac{x}{r^3} \right)\right) - \frac{\delta_{\rho\sigma}}{r}\right].
\end{equation}

\noindent In the equation, the second and fourth terms cancel out because \(\delta_{\rho\sigma}\) is 1 when \(\rho = \sigma\) and 0 otherwise. Therefore, we have:
\begin{equation}
-\delta_{\rho\sigma} + \alpha\left(x \cdot \left( -\frac{x}{r^3} \right)\right).
\end{equation}

\noindent This leads us to Einstein's spatial part in the first approximation, equation \eqref{Eq 561}.

Karl Schwarzschild, who was leading the Astrophysical Observatory in Potsdam, was the first to present an exact solution to Einstein's vacuum field equations, which Einstein had introduced in his November 18, 1915, paper, "Explanation of the Perihelion Motion of Mercury from the General Theory of Relativity" \cite{Einstein15}.

Schwarzschild, while stationed at the Russian front, wrote to Einstein on December 22, 1915, indicating his intention to refine Einstein's earlier calculations on Mercury's perihelion movement, which Einstein had presented on November 18, 1915, as evidence of the accuracy of his general theory of relativity. 

In response to Einstein's first-order approximation solution \eqref{Eq 29}, Schwarzschild discovered another similar solution \cite{CPAE8}, Doc. 169:

\begin{equation} \label{Eq 501}
g_{\rho\sigma} = -\frac{\alpha (x_{\rho} x_{\sigma})}{r^3} - \frac{\beta (x_{\rho} x_{\sigma})}{r^5} + \delta_{\rho\sigma} \left[ \frac{\beta}{3r^3} \right], \quad g_{44} = 1
\end{equation}

\noindent Like the previous term, the second term involves a constant $\beta$ and the product of the coordinate functions. However, it is divided by $r^5$, indicating a different scaling with distance. This term represents a higher-order correction. The third term is the Kronecker delta, which contributes to the diagonal elements of the metric tensor and scales with $\frac{\beta}{3r^3}$. It modifies the metric's scaling properties. $g_{44} = 1$ states that the time-time component of the metric tensor is equal to $1$.   

Following this, Schwarzschild excluded the first term (setting $\alpha =0$) in equation \eqref{Eq 501} and presented his approximate solution that closely resembled the form of Einstein's solution:

\begin{equation}
g_{\rho\sigma} = -\frac{\beta (x_{\rho} x_{\sigma})}{r^5} + \delta_{\rho\sigma} \left[\frac{\beta}{3r^3}\right], \quad g_{44} = 1.
\end{equation}

Schwarzschild aimed to show that if there are multiple approximate solutions, the problem becomes physically indeterminate.
Furthermore, equations \eqref{Eq 29} and equation \eqref{Eq 501} mix elements characteristic of both spherical and Cartesian coordinates. 

Subsequently, Schwarzschild transitioned to addressing the complete solution. He realized that, unlike the first-order approximate solutions, only one specific line element met all of Einstein's criteria for the Sun's gravitational field—static, spherically symmetric, satisfying certain metric conditions, and reverting to a flat metric at great distances (defined in section \ref{2})—as well as the vacuum field equations \eqref{Eq 30} from Einstein's November 18, 1915 paper. 

Schwarzschild conveyed the following line element to Einstein in his correspondence \cite{CPAE8}, Doc. 169:

\begin{equation} \label{Eq 500-1}
ds^2 = \left(1 - \frac{\gamma}{R}\right) dt^2 - \frac{dR^2}{1 - \frac{\gamma}{R}} - R^2 \left(d\theta^2 + \sin^2\theta \, d\phi^2\right).
\end{equation}

The symmetry and static nature of the solution meant that the only independent variable was the radial distance from the center. Near the origin ($R = 0$), Schwarzschild's solution exhibited a mathematical singularity that could not be eliminated through coordinate transformations, marking a point where spacetime dynamics cease.
Schwarzschild clarified to Einstein that his solution had just one true singularity at $R = 0$. 
He further pointed out that his spherical coordinates are not "admissible," i.e., do not conform to the requirements of Einstein's field equations \eqref{Eq 30}. 
Indeed, these coordinates do not align with Einstein's condition for unimodular coordinates, where the square root of the negative determinant is equal to $1$ [equation \eqref{Eq 529}].

Schwarzschild stressed that the spherical coordinates nonetheless provide the most suitable representation for spherical symmetry. 
This was evident as Schwarzschild's solution precisely matched Einstein's first-order approximation for Mercury's orbit \eqref{Eq 68} (as explored in detail in section \ref{6}) \cite{CPAE8}, Doc. 169. 

On December 29, 1915, Einstein responded by recognizing Schwarzschild's solution as uniquely significant \cite{CPAE8}, Doc. 176. Schwarzschild further communicated this solution to Einstein in a manuscript, deriving it from Einstein's November 18, 1915, field equations for a single mass. At the start of January 1916, Einstein received and reviewed Schwarzschild's manuscript keenly. He was pleasantly surprised by Schwarzschild's ability to formulate a precise solution to the problem, something he had not anticipated to be so straightforward. In mid-January 1916, Einstein presented Schwarzschild's paper at the Prussian Academy, offering a brief explanation to accompany it (as mentioned in a letter from Einstein to Schwarzschild on January 9, 1916) \cite{CPAE8}, Doc. 181. Schwarzschild's paper, "On the Gravitational Field of a Point-Mass according to Einstein's Theory," was published a month later \cite{Schwarzschild}.

\section{Meeting the criteria set by Einstein} \label{2}

In his paper, Schwarzschild first considered Einstein's geodesic equation \cite{Einstein1} \cite{Schwarzschild}:

\begin{equation} \label{Eq 33}
\frac{d^2 x_\tau}{ds^2} = \Gamma_{\mu\nu}^\tau \frac{dx_\mu}{ds} \frac{dx_\nu}{ds},
\end{equation}

\noindent where:

\begin{equation} \label{Eq 278}
\Gamma_{il}^\kappa = \frac{1}{2} g^{\kappa\lambda} \left( \frac{\partial g_{\lambda l}}{\partial x^i} + \frac{\partial g_{i\lambda}}{\partial x^l} - \frac{\partial g_{il}}{\partial x^\lambda} \right).
\end{equation}

\noindent are the components of the gravitational field (the Christoffel symbols). 

He then considered Einstein's vacuum field equations from his November 18, 1915, paper \cite{Einstein15}, \cite{Schwarzschild}: 

\begin{equation} \label{Eq 30}
\sum_{\alpha} \frac{\partial \Gamma_{\mu\nu}^{\sigma}}{\partial x_{\alpha}} + \sum_{\alpha\beta} \Gamma_{\mu\beta}^{\sigma} \Gamma_{\nu\alpha}^{\beta} = 0.
\end{equation}

Einstein imposed the determinant condition from his November 11, 1915 paper, "On the General Theory of Relativity (Addendum)" \cite{Einstein6}. which implies that the determinant of the metric tensor equals $-1$ \cite{Einstein15}:

\begin{equation} \label{Eq 527}
|g_{\mu\nu}| = -1.    
\end{equation}

\noindent $|g_{\mu\nu}|$ represents the absolute value of the determinant of the metric tensor.
If the determinant of the metric tensor $g_{\mu\nu}$ is equal to $-1$, then taking the square root of the negative determinant would give $1$. This can be expressed as: 

\begin{equation}
\sqrt{|g_{\mu\nu}|} = \sqrt{-(-1)} = 1.  
\end{equation}

\noindent Equation \eqref{Eq 527} represents the condition for unimodular coordinates:

\begin{equation} \label{Eq 529}
    \sqrt{-g} = 1.
\end{equation}

The condition \eqref{Eq 529} was essential for Einstein's calculation of Mercury’s perihelion advance. When Einstein worked on explaining the perihelion of Mercury, he was dealing with a very specific and complex problem. The orbit of Mercury deviates slightly from what Newtonian mechanics predicts, a discrepancy that needs a precise explanation. In this context, Einstein used the unimodular condition \eqref{Eq 529} as part of an approximate method to simplify the problem. This approach allowed him to derive a solution that could explain the observed perihelion shift without delving into a more comprehensive solution. By focusing on an approximate solution, Einstein acknowledged that his approach was a stepping stone toward a more complete understanding. Einstein clarified: "Following this conviction, I shall confine myself, for the time being, here to derive a solution, without entering into the issue of whether it is the only possible one" \cite{Einstein15}. By not questioning the uniqueness of his solution at that time, Einstein was accepting that his explanation, though valid and successful in explaining Mercury's orbit, might be one of several possible solutions within the broader framework of his general relativity. 

Recall from section \ref{1} that Einstein incorporated a combination of Cartesian coordinates and spherical coordinates in his initial approximate solution. Einstein used a hybrid approach and mixture to simplify his calculations and address the calculation of the precession of Mercury's perihelion. Einstein's approximation uses a mix of coordinate systems to highlight certain aspects of the physics involved.

Schwarzschild's contribution can be seen as a resolution to the hybrid coordinate approach utilized by Einstein, which combined Cartesian and spherical coordinate elements. Schwarzschild demonstrated remarkable ingenuity in developing a solution that not only adhered to Einstein's field equations under the unimodular restriction but also satisfied the additional conditions set forth by Einstein, conditions $1-4$ specified below. This development represented a significant advancement in general relativity, demonstrating Schwarzschild's ability to navigate and reconcile the complexities of Einstein's theoretical framework. His work provided a more precise, unified, and consistent approach than Einstein's. 

Unlike Einstein, Schwarzschild sought an exact solution to Einstein's field equations that satisfied the four criteria outlined in Einstein's paper on Mercury's perihelion \cite{Einstein15}, \cite{Schwarzschild}:
\vspace{1mm} %1mm vertical space

1) All components of $g_{\mu \nu}$ are independent of $x_4$ (time independence). This implies a static field.

2) $g_{\rho 4} = g_{4 \rho} = 0$ for $\rho = 1,2,3$. This criterion simplifies the metric by eliminating terms that mix spatial and temporal components.

3) Spherical symmetry. This is crucial for deriving Schwarzschild's solution.

4) The solution vanishes at infinity. This ensures that the spacetime metric approaches the flat Minkowski metric at great distances from the mass. 

\vspace{1mm} %1mm vertical space

According to Schwarzschild, the challenge lies in identifying a line element with coefficients that simultaneously satisfy the field equations \eqref{Eq 30}, the determinant condition \eqref{Eq 527}, and the four requirements \cite{Schwarzschild}.

\section{Schwarzschild's trick for addressing the determinant condition} \label{3}

Schwarzschild first expressed the most general line element satisfying the four conditions in terms of rectangular coordinates ($x, y, z$) and time ($t$): 

\begin{equation} \label{Eq 505}
ds^2 = F dt^2 - G(dx^2 + dy^2 + dz^2) - H(xdx + ydy + zdz)^2.
\end{equation}

The line element $ds^2$ is described using functions $F$, $G$, and $H$, which depend on the radial distance $r$ defined by equation \eqref{Eq 564}. At $r\to \infty$, the conditions $F = G = 1$ and $H = 0$ are imposed. This aligns with condition 4), i.e., with the expectation that the spacetime should resemble flat Minkowski spacetime at great distances \cite{Schwarzschild}. 

Schwarzschild then changes coordinates from rectangular to polar coordinates ($r, \theta, \phi $) according to 
\begin{align*}
x &= r \sin \theta \cos \phi, \\
y &= r \sin \theta \sin \phi, \\
z &= r \cos \theta,
\end{align*}

\noindent Schwarzschild's line element \eqref{Eq 505} in polar coordinates reads:

\begin{equation} \label{Eq 506}
ds^2 = F dt^2 - G \left( dr^2 + r^2 d\theta^2 + r^2 \sin^2\theta d\phi^2 \right) - H r^2 dr^2,
\end{equation}

\noindent which is equivalent to: 

\begin{equation} \label{Eq 544}
ds^2 = F dt^2 - (G + Hr^2) dr^2 - Gr^2 (d\theta^2 + \sin^2\theta d\phi^2).
\end{equation}

\noindent $ds^2$ represents the square of the differential element of spacetime distance, $dt^2$, $dr^2$, $d\theta^2$, and $d\phi^2$ are the differential elements of time, radial distance, polar angle, and azimuthal angle, respectively. $F, G, H$ are functions of $r$ \cite{Schwarzschild}.

When transforming the line element \eqref{Eq 505} to polar coordinates, Schwarzschild faced the issue that the transformed volume element in polar coordinates: 

\begin{equation} \label{Eq 543}
r^2 \sin \theta \, dr \, d\theta \, d\phi, 
\end{equation}

\noindent does not maintain the determinant condition of $1$ [equation \eqref{Eq 527}]. 

Schwarzschild, while at the Russian front, ingeniously resolved this issue through a clever approach, demonstrating his exceptional mathematical skill. 
This approach highlighted that the apparent problem in Schwarzschild's solution was not inherent to the proposed metric but rather a consequence of the spherical coordinate system he used.  This solution, as presented in Schwarzschild's 1916 paper \cite{Schwarzschild}, is a testament to his mathematical prowess and ability to navigate complex theoretical issues. 

Specifically, Schwarzschild introduced a coordinate transformation, i.e., new variables $x_1, x_2, x_3$ \cite{Schwarzschild}: 

\begin{equation} \label{Eq 507}
x_1 = \frac{r^3}{3}, x_2 = -\cos \theta, x_3 = \phi, x_4 = t,
\end{equation}

\noindent which he called: "polar coordinates with the determinant $1$" \cite{Schwarzschild}.

The differentials of these new coordinates in terms of the old ones are:
\begin{equation}
dx_1 = r^2 \, dr, \quad dx_2 = \sin \theta \, d\theta, \quad dx_3 = d\phi.
\end{equation}

\noindent Substituting these into the volume element in spherical coordinates \eqref{Eq 543}, we get:

\begin{equation} \label{Eq 531}
dV = dx_1 \, dx_2 \, dx_3
\end{equation}

\noindent Schwarzschild's "trick," i.e., a clever and non-obvious method used to simplify and solve the complex problem of Einstein's determinant condition \eqref{Eq 527} in the form of coordinate transformations, transforms the volume element in spherical coordinates \eqref{Eq 543} to equation \eqref{Eq 531}.

Schwarzschild emphasizes that the new variables retain the advantages of polar coordinates for solving the problem (since they are modified polar coordinates). Schwarzschild stresses that the altered radial and angular relationships in equation \eqref{Eq 507} simplify calculations but do not alter the form of the field equations \eqref{Eq 30} or the determinant condition \eqref{Eq 527}. By including time as $t=x_4$, Schwarzschild ensures that both the field equations and the determinant equation of the metric tensor remain unaltered in form.

Schwarzschild then rewrote the line element \eqref{Eq 544} using the new variables \eqref{Eq 507} \cite{Schwarzschild}:

\begin{equation} \label{Eq 502}
ds^2 = F dx_4^2 - \left(\frac{G}{r^4}+\frac{H}{r^2}\right) dx_1^2 - Gr^2 \left[\frac{ dx_2^2}{1- dx_2^2}+ dx_3^2(1- dx_2^2)\right],
\end{equation}

\noindent and with four functions $f_1, f_2, f_3, f_4$ \cite{Schwarzschild}:

\begin{equation} \label{Eq 530}
ds^2 = f_4 \, dx_4^2 - f_1 \, dx_1^2 - f_2 \,\frac{dx_2^2}{1 - x_2^2} - f_3 \, dx_3^2 \, (1 - x_2^2), \quad \text{  where: }
\end{equation}

\begin{equation} \label{Eq 517}
g_{11} = -f_1, g_{22} = -\frac{f_2}{1 - x_2^2},  g_{33} = -f_2({1 - x_2^2}), g_{44} = f_4. 
\end{equation}

Schwarzschild derived three (four, due to one redundancy) functions $f_1, f_2, f_3, f_4$ that satisfied Einstein's conditions and the vacuum field equations \eqref{Eq 30}. These equations are simplified and constrained by the determinant condition, which affects the metric components. 

Schwarzschild first writes the differential equations of the geodesic line for the line element \eqref{Eq 530}. These equations are obtained using the variational principle, leading to expressions in terms of the second derivative of the coordinates with respect to an affine parameter $s$ and partial derivatives of the metric functions $f_1, f_2$, and $f_4$ with respect to $x_1$ \cite{Schwarzschild}:

\begin{equation} \label{Eq 571a}
0 = f_1 \frac{d^2 x_1}{ds^2} + \frac{1}{2} \frac{\partial f_4}{\partial x_1} \left(\frac{dx_4}{ds}\right)^2 + \frac{1}{2} \frac{\partial f_1}{\partial x_1} \left(\frac{dx_1}{ds}\right)^2 \end{equation}

\begin{equation*}
- \frac{1}{2} \left[\frac{\partial f_2}{\partial x_1} \frac{1}{1 - x_2^2} \left(\frac{dx_2}{ds}\right)^2 + (1-x^2_2)\left(\frac{dx_3}{ds}\right)^2\right],    
\end{equation*}

\begin{equation} \label{Eq 571b}
0 = \frac{f_2}{1 - x_2^2} \frac{d^2 x_2}{ds^2} + \frac{\partial f_2}{\partial x_1} \frac{1}{1 - x_2^2} \frac{dx_1}{ds} \frac{dx_2}{ds}
\end{equation}

\begin{equation*}
+ \frac{f_2 x_2}{(1 - x_2^2)^2} \left(\frac{dx_2}{ds}\right)^2 + f_2 x_2 \left(\frac{dx_3}{ds}\right)^2,    
\end{equation*}

\begin{equation} \label{Eq 571c}
0 = f_2 (1 - x_2^2) \frac{d^2 x_3}{ds^2} + \frac{\partial f_2}{\partial x_1} (1 - x_2^2) \frac{dx_1}{ds} \frac{dx_3}{ds} - 2f_2 x_2 \frac{dx_2}{ds} \frac{dx_3}{ds},
\end{equation}

\begin{equation} \label{Eq 571d}
0 = f_4 \frac{d^2 x_4}{ds^2} + \frac{\partial f_4}{\partial x_1} \frac{dx_1}{ds} \frac{dx_4}{ds}.
\end{equation}

The gravitational field components (Christoffel symbols) are derived by matching the specific geodesic equations \eqref{Eq 571a}, \eqref{Eq 571b}, \eqref{Eq 571c}, and \eqref{Eq 571d} with the general geodesic equation \eqref{Eq 33}. 
Each term in the specific geodesic equations corresponds to a particular Christoffel symbol. 
\begin{equation*}
\text{For instance, the term } \frac{1}{2}\frac{\partial f_4}{\partial x_1}\left(\frac{dx_4}{ds}\right)^2 \text{ in equation \eqref{Eq 571a} contributes to } \Gamma^1_{44}.      
\end{equation*}
\noindent So, comparing equations \eqref{Eq 571a}, \eqref{Eq 571b}, \eqref{Eq 571c}, and \eqref{Eq 571d} with the general geodesic equation \eqref{Eq 33}, Schwarzschild identified the components of the gravitational field \cite{Schwarzschild}:

\begin{equation} \label{Eq 572a}
\begin{aligned}
\Gamma^1_{11} = -\frac{1}{2} \frac{1}{f_1} \frac{\partial f_1}{\partial x_1}, \\ \Gamma^1_{22} = \frac{1}{2} \frac{1}{f_1} \frac{\partial f_2}{\partial x_1} \frac{1}{1 - x_2^2}, 
\end{aligned}
\end{equation}

\begin{equation*}
\begin{aligned}
\Gamma^1_{33} = \frac{1}{2} \frac{1}{f_1} \frac{\partial f_2}{\partial x_1} (1 - x_2^2), \\ \Gamma^1_{44} = -\frac{1}{2} \frac{1}{f_1} \frac{\partial f_4}{\partial x_1}.    
\end{aligned}
\end{equation*}

\begin{equation} \label{Eq 572b}
\begin{aligned}
\Gamma^2_{21} = -\frac{1}{2} \frac{1}{f_2} \frac{\partial f_2}{\partial x_1}, \\ \Gamma^2_{22} = -\frac{x_2}{1 - x_2^2}, \\ \Gamma^2_{33} = -x_2 (1 - x_2^2), 
\end{aligned}
\end{equation}

\begin{equation*}
\begin{aligned}
\Gamma^3_{31} = -\frac{1}{2} \frac{1}{f_2} \frac{\partial f_2}{\partial x_1}, \\ \Gamma^3_{32} = \frac{x_2}{1 - x_2^2}.    
\end{aligned}
\end{equation*}

\begin{equation} \label{Eq 572c}
\Gamma^4_{41} = -\frac{1}{2} \frac{1}{f_4} \frac{\partial f_4}{\partial x_1}.
\end{equation}

Equations \eqref{Eq 572a}, \eqref{Eq 572b}, and \eqref{Eq 572c} involve only derivatives of the functions $f_1, f_2, f_3, f_4$ with respect to $x_1$, simplifying the computation of the field equations.

Schwarzschild then refers to simplifying the field equations \eqref{Eq 30} due to the rotational symmetry around the origin, specifically for the equator where $x_2 =0$. 
In such a scenario, we can simplify the expressions by setting $1-x^2_2=1$ from the beginning, significantly simplifying the differential equations \eqref{Eq 572a}, \eqref{Eq 572b}, and \eqref{Eq 572c}. 

So, given the rotational symmetry and focusing on the equator, all terms involving $x_2$ disappear, and terms in equations \eqref{Eq 572a}, \eqref{Eq 572b}, and \eqref{Eq 572c} involving $1-x^2_2$ can be simplified since $1-x^2_2=1$ \cite{Schwarzschild}: 

\begin{equation} \label{Eq 573a}
\frac{\partial}{\partial x_1} \left( \frac{1}{f_1} \frac{\partial f_1}{\partial x_1} \right) = \frac{1}{2} \left( \frac{1}{f_1} \frac{\partial f_1}{\partial x_1} \right)^2 + \left( \frac{1}{f_2} \frac{\partial f_2}{\partial x_1} \right)^2 + \frac{1}{2} \left( \frac{1}{f_4} \frac{\partial f_4}{\partial x_1} \right)^2,
\end{equation}

\begin{equation} \label{Eq 573b}
\frac{\partial}{\partial x_1} \left( \frac{1}{f_1} \frac{\partial f_2}{\partial x_1} \right) = 2 + \frac{1}{f_1 f_2} \left( \frac{\partial f_2}{\partial x_1} \right)^2,
\end{equation}

\begin{equation} \label{Eq 573c}
\frac{\partial}{\partial x_1} \left( \frac{1}{f_1} \frac{\partial f_4}{\partial x_1} \right) = \frac{1}{f_1 f_4} \left( \frac{\partial f_4}{\partial x_1} \right)^2.
\end{equation}

Given the line element \eqref{Eq 530}, the determinant of the metric tensor depends on the functions $f_1, f_2, f_4$. The unimodular coordinate condition \eqref{Eq 527} hence translates into a specific relationship among these functions. The equation, as per the condition, can be written as \cite{Schwarzschild}:

\begin{equation} \label{Eq 574}
f_1 f_2^2 f_4 = 1.
\end{equation}

\noindent We calculate the determinant of $g_{\mu \nu}$ using the equation \eqref{Eq 517}: 

\begin{equation}
g_{\mu\nu} =\begin{pmatrix}
-f_1 & 0 & 0 & 0 \\
0 & -\frac{f_2}{1 - x_2^2} & 0 & 0 \\
0 & 0 & -f_2(1 - x_2^2) & 0 \\
0 & 0 & 0 & f_4.
\end{pmatrix}
\end{equation}

\noindent The determinant of the metric tensor is the product of its diagonal elements:

\begin{equation}
|g_{\mu\nu}| = (-f_1) \times \left(-\frac{f_2}{1 - x_2^2}\right) \times \left(-f_2(1 - x_2^2)\right) \times f_4,
\end{equation}

\noindent from which we get: 

\begin{equation}
|g_{\mu\nu}| = -f_1 f_2^2 f_4 = -1.
\end{equation}

\noindent Notice that the actual determinant, considering the signature of the metric, is negative. In Schwarzschild's expression \eqref{Eq 574}, the absolute value of the determinant of the metric tensor is $1$. However, the actual determinant could be either $+1$ or $-1$, depending on the signature of the metric. 
Equation \eqref{Eq 574} implies \cite{Schwarzschild}:

\begin{equation} \label{Eq 575}
\frac{1}{f_1} \frac{\partial f_1}{\partial x_1} + \frac{2}{f_2} \frac{\partial f_2}{\partial x_1} + \frac{1}{f_4} \frac{\partial f_4}{\partial x_1} = 0.
\end{equation}

Equation \eqref{Eq 575} balances the derivatives of the metric functions $f_1, f_2, f_4$ to maintain the metric tensor's determinant as a constant (i.e., unity). It is a constraint that these functions must satisfy in addition to the field equations \eqref{Eq 573a}, \eqref{Eq 573b}, \eqref{Eq 573c}. The equation \eqref{Eq 575} guarantees that the rate of change of $f_1, f_2, f_4$ in the coordinates of line element \eqref{Eq 530} is such that their combined effect keeps the determinant of the metric tensor equal to one, equation \eqref{Eq 574}.

For now, Schwarzschild neglects the field equation \eqref{Eq 573b}, and rewrites the field equation involving $f_4$, equation \eqref{Eq 573c} \cite{Schwarzschild}:

\begin{equation} \label{Eq 576}
\frac{\partial}{\partial x_1} \left( \frac{1}{f_4} \frac{\partial f_4}{\partial x_1} \right) = \frac{1}{f_1 f_4} \frac{\partial f_1}{\partial x_1} \frac{\partial f_4}{\partial x_1}.
\end{equation}

\noindent He directly integrates equation \eqref{Eq 576} to yield a relationship between the derivatives of $f_1$ and $f_4$. This results in:

\begin{equation} \label{Eq 578}
\frac{1}{f_4} \frac{\partial f_4}{\partial x_1} = \alpha f_1 \text{ 
 where } \alpha \text{ is an integration constant.}
\end{equation}

\noindent By combining the field equation for $f_1$ \eqref{Eq 573a} with the transposed equation \eqref{Eq 576} for $f_4$, Schwarzschild got another equation relating the derivatives of $f_1, f_2$, and $f_4$: 

\begin{equation}
\frac{\partial}{\partial x_1} \left( \frac{1}{f_1} \frac{\partial f_1}{\partial x_1} + \frac{1}{f_4} \frac{\partial f_4}{\partial x_1} \right) = \left( \frac{1}{f_2} \frac{\partial f_2}{\partial x_1} \right)^2 + \frac{1}{2} \left( \frac{1}{f_1} \frac{\partial f_1}{\partial x_1} + \frac{1}{f_4} \frac{\partial f_4}{\partial x_1} \right)^2.
\end{equation}

\noindent This combination leads to an equation involving only the derivative of $f_2$. In other words, using the determinant condition \eqref{Eq 575}, Schwarzschild eliminated terms involving the derivatives of $f_1$ and $f_4$ to get the equation that involves only the derivative of $f_2$:

\begin{equation} \label{Eq 577}
-2 \frac{\partial}{\partial x_1} \left( \frac{1}{f_2} \frac{\partial f_2}{\partial x_1} \right) = 3 \left( \frac{1}{f_2} \frac{\partial f_2}{\partial x_1} \right)^2.
\end{equation}

$\textsc{Equation for } f_2$:
\vspace{1mm} %1mm vertical space

\noindent Next, Schwarzschild integrates the equation \eqref{Eq 577} for $f_2$:

\begin{equation} \label{Eq 579}
\frac{1}{\frac{1}{f_2} \frac{\partial f_2}{\partial x_1}} = \frac{3}{2} x_1 + \frac{\rho}{2} \text{ Alternate form of the equation: } \frac{1}{f_2} \frac{\partial f_2}{\partial x_1} = \frac{2}{3x_1 + \rho}.
\end{equation}

\noindent where $\rho$ is a constant of integration.

The condition at infinity [see equations \eqref{Eq 583}, \eqref{Eq 584}, \eqref{Eq 585} further below] is used to set integration constants, ensuring that the function $f_2$ approaches expected values as $x_1$ goes to infinity. 

\noindent Hence, Schwarzschild integrates equation \eqref{Eq 579} and gets:

\begin{equation} \label{Eq 580}
f_2 = \lambda(3x_1 + \rho)^{2/3}  \quad \text{ where: } \lambda \text{ is an integration constant.}
\end{equation}

\noindent The condition at infinity demands: $\lambda =1$, and equation \eqref{Eq 580} gives $f_2$:

\begin{equation} \label{Eq 504a}
f_2 = (3x_1 + \rho)^{\frac{2}{3}}.    
\end{equation}

$\textsc{Equation for } f_4$:
\vspace{1mm} %1mm vertical space

Using the equation \eqref{Eq 578} for $f_4$ and the determinant condition \eqref{Eq 575} we find an expression for $\frac{\partial f_4}{\partial x_1}$ in terms of $f_1, f_2, f_4$
The expression becomes:

\begin{equation} \label{Eq 581}
\frac{\partial f_4}{\partial x_1} = \alpha f_1 f_4 = \alpha \left( \frac{1}{f_2} \right)^2 = \frac{\alpha}{\left( 3x_1 + \rho \right)^{4/3}}.
\end{equation}

\noindent Integrating this with respect to $x_1$ and applying the condition at infinity gives the final form of $f_4$:

\begin{equation} \label{Eq 504b}
f_4 = 1 - \alpha(3x_1 + \rho)^{-\frac{1}{3}}.    
\end{equation}

$\textsc{Equation for } f_1$:
\vspace{1mm} %1mm vertical space

Schwarzschild then substituted the expressions \eqref{Eq 504a} and \eqref{Eq 504b} into the determinant condition \eqref{Eq 575}. The terms involving $f_2$ and $f_4$ are now known functions of $x_1$. The remaining term, involving $f_1$ must be such that the entire expression sums to zero. The condition \eqref{Eq 575} forms a differential equation for $f_1$ in terms of $x_1$. By integrating this differential equation, we get an expression for $f_1$. Schwarzschild's main scheme involves expressing $\frac{\partial f_2}{\partial x_1}$ and $\frac{\partial f_4}{\partial x_1}$ in terms of $x_1$. Rearranging the determinant condition to isolate the term involving $\frac{\partial f_1}{\partial x_1}$, and integrating this term to find $f_1$. The result is: 

\begin{equation} \label{Eq 504c}
f_1 = \frac{(3x_1 + \rho)^{-\frac{4}{3}}}{1 - \alpha(3x_1 + \rho)^{-\frac{1}{3}}}.     
\end{equation}

Finally, recall that Schwarzschild neglected the field equation \eqref{Eq 573b}. He notes that the two equations \eqref{Eq 504a} and \eqref{Eq 504c} inherently fulfill this equation. From equation \eqref{Eq 504a}, we compute the derivative of $f_2$ with respect to $x_1$. We plug this derivative into the field equation \eqref{Eq 573b}. With $f_1$ defined in equation \eqref{Eq 504c}, we substitute it into the field equation \eqref{Eq 573b}, and finally show that the left-hand side of the field equation \eqref{Eq 573b} simplifies to the right-hand side. 

So, Schwarzschild found from equations \eqref{Eq 580} and \eqref{Eq 581} expressions for $f_1, f_2, f_4$. Combining equations \eqref{Eq 504a}, \eqref{Eq 504b}, and \eqref{Eq 504c} gives \cite{Schwarzschild}:

\begin{equation} \label{Eq 504}
\begin{aligned}
f_2 &= f_3 = (3x_1 + \rho)^{\frac{2}{3}}, \\
f_4 &= 1 - \alpha(3x_1 + \rho)^{-\frac{1}{3}}, \\
f_1 &= \frac{(3x_1 + \rho)^{-\frac{4}{3}}}{1 - \alpha(3x_1 + \rho)^{-\frac{1}{3}}}. 
\end{aligned}
\end{equation}
\vspace{1mm} %1mm vertical space

\noindent The following criteria guarantee the validity of equations \eqref{Eq 504} \cite{Schwarzschild}: 
\vspace{1mm} %1mm vertical space

1. \emph{The condition at infinity}: The metric functions $f_1, f_2, f_3$, and $f_4$ are functions of $x_1$. Schwarzschild assumes a scaling of the spacetime metric components with respect to $x_1$ in his model. 

For $x_1 = \infty$:

\begin{equation} \label{Eq 583}
f_1 = \frac{1}{r^4} = (3x_1)^{-\frac{4}{3}}.
\end{equation}

\noindent Hence, the influence of the metric component $f_1$ diminishes at large distances from the gravitational source. This behavior is consistent with Einstein's condition of asymptotic flatness, where spacetime becomes increasingly flat as one moves away from the gravitational source.
The condition equation \eqref{Eq 583} sets $f_1$ as inversely proportional to the fourth power of the radius $r$ defined by equation \eqref{Eq 564}. 

Next, the metric functions $f_2=f_3$, for $x_1 = \infty$:

\begin{equation} \label{Eq 584}
f_2 = f_3 = r^2 = (3x_1)^{\frac{2}{3}}.
\end{equation}

\noindent The expression \eqref{Eq 584} shows that both $f_2$ and $f_3$ increase as $x_1$ increases. In other words, as $x_1$ approaches infinity, both $f_2$ and $f_3$ will grow without bounds. This suggests that the spatial components of the metric associated with $f_2$ and $f_3$ expand as one moves further away from the central region. The specific power of $\frac{2}{3}$ indicates how the spatial dimensions expand as a function of $x_1$. 

While $f_1$ decreases to zero at infinity, $f_2$ and $f_3$ increase. The asymptotic behavior of these metric components is consistent with a spacetime that becomes flat at large distances from a spherically symmetric mass.  

Finally, for $f_4$: 

\begin{equation} \label{Eq 585}
f_4 = 1.
\end{equation}

\noindent This condition sets $f_4$ as 1 when $x_1 = \infty$. In other words, as one moves away from the gravitational source to infinity, the spacetime behaves more like flat spacetime.

2. \emph{The determinant condition}: After transforming to the new variables \eqref{Eq 507} and defining the functions: 

\begin{equation*}
f_1, f_2, f_3, f_4,    
\end{equation*}

\noindent Schwarzschild needed to ensure that the product of these functions:

\begin{equation}
\sqrt{-g} = \sqrt{f_1 \times f_2 \times f_3 \times f_4},    
\end{equation}

\noindent equaled $1$, thereby satisfying the determinant condition \eqref{Eq 527}, essential for simplifying and solving Einstein's vacuum field equations \eqref{Eq 30}.

To fulfill this requirement, Schwarzschild specifically chose and determined the functions such that their product, when considered with respect to the metric tensor for the line element \eqref{Eq 502} and its determinant, resulted in $-1$. This was a crucial step in Schwarzschild's derivation, as it allowed him to successfully apply Einstein's field equations \eqref{Eq 30} to find his exact solution. 

3. \emph{The continuity condition}: ensuring that the functions $f_1, f_2, f_3, f_4$ are continuous, except at $x_1 = 0$.

Schwarzschild realized that the functions $f_1, f_2, f_3, f_4$ violate the continuity condition. He identified a mathematical singularity in the expression for $f_1$. If:  

\begin{equation} \label{Eq 566}
\alpha(3x_1 + \rho)^{-\frac{1}{3}} = 1, \text{ and }      
\end{equation}

\begin{equation} \label{Eq 567}
3x_1 = \alpha^3 - \rho, \text{ then } f_1 \text{ is singular.}    
\end{equation}

\noindent Let us do the simple calculation by plugging the second condition equation \eqref{Eq 567}:

\begin{equation} \label{Eq 568}
3x_1 + \rho = \alpha^3,    
\end{equation}

\noindent into the first one equation \eqref{Eq 566}. We get: 

\begin{equation} \label{Eq 569}
\alpha \cdot \alpha ^{-1} = 1.  
\end{equation}

\noindent Thus, the two conditions \eqref{Eq 566} and \eqref{Eq 567} are consistent, confirming that the denominator of $f_1$ equation \eqref{Eq 504} is zero under these conditions, making the function singular.
Let us confirm this by substituting equation \eqref{Eq 568} into the expression for $f_1$ equation \eqref{Eq 504}. The simplified expression for $f_1$ under the conditions \eqref{Eq 566} and \eqref{Eq 567} becomes:

\begin{equation} \label{Eq 570}
f_1 = \frac{(\alpha^3)^-{\frac{4}{3}}}{1- \alpha \cdot (\alpha ^3){-\frac{1}{3}}}=\frac{\alpha^{-4}}{1-\alpha \cdot \alpha ^{-1}}=\frac{0}{\alpha^4}.     
\end{equation}

\noindent The function becomes undefined and singular. 

To explore the discontinuity coinciding with the origin, Schwarzschild set $x_1=0$ in equation \eqref{Eq 568} and obtained:

\begin{equation} \label{Eq 528}
\rho = \alpha^3.
\end{equation}

\noindent Schwarzschild redefined the function $f_1$ by relating the integration constants $\alpha$ and $\rho$\cite{Schwarzschild}. In other words, Schwarzschild finds that the continuity condition at the origin relates the two integration constants in the specific way of \eqref{Eq 528}. This condition ensures that any potential discontinuity in the solution is located at the origin, a requirement for physical validity.
Thus, evaluating at the origin (where $x_1=0$) reveals that the discontinuity coincides with the origin:

\begin{equation} \label{Eq 588}
f_1(0)=\frac{\rho^-\frac{4}{3}}{1-\alpha \rho^ -\frac{1}{3}}    =\frac{\alpha^{-4}}{1-\alpha \cdot \alpha ^{-1}}=\frac{0}{\alpha^4}.    
\end{equation}

\noindent This discontinuity occurs regardless of the specific value of $\rho$, as long as it follows the relationship equation \eqref{Eq 528}.

\section{Schwarzschild reverts to spherical coordinates} \label{4}

After obtaining the metric functions -- equations \eqref{Eq 504a}, \eqref{Eq 504b}, and \eqref{Eq 504c}  [equation \eqref{Eq 504}]-- satisfying the determinant condition \eqref{Eq 574}, Schwarzschild rewrote them. 
With the relationship \eqref{Eq 528} established, he reformulated the functions $f_1, f_2, f_3$ and $f_4$ in terms of a new transformed radial coordinate $R$ \cite{Schwarzschild}. 

First, Schwarzschild used the transformation equation \eqref{Eq 507} to rewrite the condition equation \eqref{Eq 568}. This led to the formulation of the Schwarzschild radius $R$ equation: 

\begin{equation} \label{Eq 536}
R = \left(3x_1 + \rho\right)^{\frac{1}{3}} = \left(r^3 + \alpha^3\right)^{\frac{1}{3}},    
\end{equation}

\noindent where $r$ represents the standard radial coordinate in spherical coordinates, derived from Cartesian coordinates, defined by equation \eqref{Eq 564}. 

Recall that Schwarzschild introduced the transformation equation \eqref{Eq 507} as a mathematical maneuver to simplify the solution of Einstein's field equations under the conditions of spherical symmetry. In his derivation of equation \eqref{Eq 536}, Schwarzschild redefined equation \eqref{Eq 568}. He substituted the transformation equation \eqref{Eq 507} into the condition equation \eqref{Eq 568} and expressed the condition in terms of the original radial coordinate $x_1 = \frac{r^3}{3}$ or $3x_1 = r^3$, thereby linking $R$ with the original radial coordinate $r$. Since the $f_1, f_2=f_3$, and $f_4$ are only functions of $x_1$, $x_2=x_3=0$.

Substituting the transformation equation \eqref{Eq 507} (for $x_1$) into equation \eqref{Eq 504}, we can express the metric functions $f_1, f_2=f_3$, and $f_4$ in terms of $r$:

\begin{equation} \label{Eq 589}
f_1 = \frac{(r^3 + \rho)^{-\frac{4}{3}}}{1 - \alpha (r^3 + \rho)^{-\frac{1}{3}}}
\end{equation}
    
\begin{equation*}
f_4 = 1 - \alpha (r^3 + \rho)^{-\frac{1}{3}}, \ f_2 = f_3 = (r^3 + \rho)^{\frac{2}{3}}.
\end{equation*}

Second, Schwarzschild substituted the expression for $R$ [equation \eqref{Eq 536}] into the functions $f_1, f_2=f_3$ and $f_4$, equation \eqref{Eq 504} or \eqref{Eq 589}, and obtained: 

\begin{equation} \label{Eq 532}
\begin{aligned}
f_1 &= \frac{1}{R^4} \cdot \frac{1}{1-\frac{\alpha}{R}}, \\
f_2 &= f_3 = R^2, \\
f_4 &= 1 - \frac{\alpha}{R}.
\end{aligned}
\end{equation}
\vspace{1mm} %1mm vertical space

Plugging these metric functions back into the line element \eqref{Eq 530} and reverting to standard spherical coordinates, Schwarzschild derived the final form of the line element, the now-famous Schwarzschild line element \cite{Schwarzschild}:

\begin{equation} \label{Eq 500}
ds^2 = (1 - \frac{\alpha}{R}) dt^2 - \frac{1}{1 - \frac{\alpha}{R}} dR^2 - R^2 (d\theta^2 + \sin^2\theta d\phi^2).
\end{equation}

This line element \eqref{Eq 500} results when the functions \eqref{Eq 532} are substituted back into the line element after transforming to the usual polar coordinates \cite{Schwarzschild}. 
The Schwarzschild line element (equation \eqref{Eq 500}) describes the spacetime interval $ds^2$ in terms of the coordinates $t, R, \theta$ and $\phi$.

This new form of the line element is a unique and exact solution to the Einstein field equations \eqref{Eq 30} for a spherically symmetric and static mass in a vacuum. Schwarzschild discusses the uniqueness of his solution to Einstein's field equations \eqref{Eq 30} and the difficulties that would arise in ascertaining this uniqueness through Einstein's approximation method \cite{Schwarzschild}. 

In Schwarzschild's exact solution, the uniqueness emerged naturally during his calculations of \eqref{Eq 589}. He did not have to impose additional conditions to ensure uniqueness; it was a direct result of his approach and the mathematical structure of the problem. Schwarzschild points out that using an approximation method (like Einstein's approach) to solve the field equations would not easily reveal the uniqueness of the solution. In approximations, one typically expands the functions \eqref{Eq 589} in a series and keeps terms up to a certain order, assuming the parameters $\alpha$ and $\rho$ are small.  
In Schwarzschild's exact solution, the continuity condition \eqref{Eq 528} played a crucial role. It linked the constants $\alpha$ and $\rho$ so that the potential discontinuity in the solution would occur at the origin (the location of the mass point), \eqref{Eq 588}. Without this condition, the solution would appear to have arbitrary constants, leading to physical indeterminacy. Recognizing the necessary link between $\alpha$ and $\rho$ would be challenging in an approximation using series expansion. The general form of the series expansion (up to second order) of $f_1$ [in equation \eqref{Eq 589}] in terms of $\alpha$ and $\rho$, which would look like this:

\begin{equation} \label{Eq 590}
f_1 \approx \frac{1}{r^4} \left[ 1 + \frac{\alpha}{r} - \frac{4}{3} \frac{\rho}{r^3} + \cdots \right].
\end{equation}

\noindent would satisfy all the conditions specified in sections \ref{2} and \ref{3} only up to a certain accuracy. 
Up to the second order in $\alpha$ and $\rho$ we first expand each term in the equation as a power series in $\alpha$ and $\rho$. Given that this is a complex expansion, equation \eqref{Eq 590} involves a Taylor series expansion of each component of the function around $\alpha = 0$ and $\rho=0$. 

Hence, the continuity condition would not seem to add new information since discontinuities naturally occur only at the origin. However, the exact solution shows that the discontinuity does not occur at the origin; it happens at $r = (\alpha^3 - \rho)^{\frac{1}{3}}$, necessitating the setting of the equation  \eqref{Eq 528} for the discontinuity to be correctly positioned at the origin, see derivation in the previous section, equations \eqref{Eq 566}, \eqref{Eq 567}, \eqref{Eq 570}, \eqref{Eq 528}, and \eqref{Eq 588}. The exact solution thus provides a deeper insight that might be missed in an approximation approach. It shows how certain parameters must be linked to ensure physical accuracy and the correct behavior of the solution, particularly regarding the location of discontinuities and the behavior near the origin. 

The exact solution equation \eqref{Eq 500} demonstrates that the constants $\alpha$ and $\rho$ are not arbitrary but are related in a specific way to ensure physical validity. The uniqueness of the solution emerges from the analysis of equations \eqref{Eq 566}, \eqref{Eq 567}, \eqref{Eq 570}, \eqref{Eq 528}, and \eqref{Eq 588}. Without the continuity condition, the solution would appear physically undetermined. However, the exact solution shows the necessity of linking $\alpha$ and $\rho$ to ensure continuity at the origin. 

For the resulting Schwarzschild metric \eqref{Eq 500}, the determinant $g$ is negative, but $\sqrt{-g}$ simplifies to $R^2\sin \theta$ rather than $1$ \cite{Schwarzschild}. 

\noindent Calculating the determinant for the Schwarzschild metric \eqref{Eq 500} involves the product of the diagonal elements of the metric tensor. The diagonal elements are: 

\begin{equation*}
\left(1 - \frac{\alpha}{R}\right), -\frac{1}{1 - \frac{\alpha}{R}}, -R^2, -R^2 \sin^2\theta.     
\end{equation*}

\noindent The determinant $g$ is the product of these diagonal elements, giving:

\begin{equation}
g = \left(1 - \frac{\alpha}{R}\right)\cdot \left(-\frac{1}{1 - \frac{\alpha}{R}}\right) \cdot (-R^2) \cdot (-R^2\sin^2\theta) =   -R^4\sin^2 \theta.   
\end{equation}

\noindent Thus, the square root of the negative determinant $\sqrt{-g}$ simplifies to:

\begin{equation}
\sqrt{-g} = \sqrt{-\left(-R^4\sin^2 \theta\right)} = R^2 \sin\theta.  
\end{equation}

\noindent Thus:\footnote{Later, it was shown that the Schwarzschild metric could be expressed in various coordinate systems: Kruskal-Szekeres, Eddington-Finkelstein, Lemaître, Gullstrand-Painlevé, Isotropic, and Novikov coordinates. However, all these coordinate systems for writing the Schwarzschild line element naturally yield determinants of the metric tensor that are not equal to $-1$.} 

\begin{equation} \label{Eq 565}
\sqrt{|\text{-}g|} = R^2 \sin\theta.    
\end{equation}

\noindent Schwarzschild shows that his line element in spherical coordinates [equation \eqref{Eq 500}] represents a valid solution to the original, i.e., the version of Einstein's field equations \eqref{Eq 30} valid only in unimodular coordinates. However, the determinant of the Schwarzschild metric \eqref{Eq 500} is not $-1$; it varies with the radial coordinate. Hence, the Schwarzschild solution is valid in the broader context of Einstein's field equations, not limited to unimodular coordinates.  

In the weak-field limit, where the gravitational field is not too strong, general relativity should be reduced to Newtonian gravity. $\alpha$ is given by:

\begin{equation} \label{Eq 73}
\alpha = \frac{\kappa M}{4\pi} = \frac{2GM}{c^2}.
\end{equation}

\noindent Here, $\kappa$ is Einstein's gravitational constant, $G$ is the Newtonian gravitational constant, $M$ is the mass of the Sun, and $c$ is the speed of light.

\noindent In modern terminology, we define $\alpha$ in equation \eqref{Eq 73} as the Schwarzschild radius \( r_s \):

\begin{equation} \label{Eq 73a}
r_s = \frac{2GM}{c^2},    
\end{equation}

\noindent and the parameter \( r_g \) is used to denote the gravitational radius, which is half of the Schwarzschild radius: 

\begin{equation} \label{Eq 609}
r_g = \frac{GM}{c^2}.    
\end{equation}

\noindent It is a measure of the strength of the gravitational field of the mass.

Substituting $\alpha = 2Gm$ into the line element \eqref{Eq 500} the Schwarzschild line element takes the known form: 

\begin{equation} \label{Eq 537}
    ds^2 = \left(1 - \frac{2Gm}{r}\right) dt^2 - \frac{dr^2}{1 - \frac{2Gm}{r}} - r^2 \left(d\theta^2 + \sin^2\theta \, d\phi^2\right), \text{ or:}
\end{equation}

\begin{equation} \label{Eq 537a}
    ds^2 = \left(1 - \frac{r_s}{r}\right) dt^2 - \frac{dr^2}{1 - \frac{r_s}{r}} - r^2 \left(d\theta^2 + \sin^2\theta \, d\phi^2\right).
\end{equation}

I will stick to the historical terminology here. The substitution of $\alpha = 2Gm$ guarantees that the metric reduces to the Newtonian form under these conditions. Specifically, in the limit as the radial coordinate $R$ becomes much larger than $\alpha$ 
the metric should approximate the metric of weak gravitational fields. 

\noindent According to the four conditions imposed by Einstein (see section \ref{2}), as $\lim_{r \to \infty}$:

\begin{equation}
    ds^2 = dt^2 - dr^2 - r^2 \left(d\theta^2 + \sin^2\theta \, d\phi^2\right),
\end{equation}

\noindent and we obtain the Minkowski line element in the spherical symmetric form. 
A mathematical singularity is seen to occur at the origin of the Schwarzschild line element \eqref{Eq 537} when $r = 0$. Coordinate transformations cannot remove this singularity. 

\section{A qualitative comparison between Droste and Schwarzschild} \label{11}

Johannes Droste, a student of Hendrik Lorentz at the University of Leiden, independently derived a solution to Einstein's field equations around the same time as Schwarzschild.
Droste's contribution is interesting because, in his work on deriving the Schwarzschild metric, Dorste uses a series of clever coordinate transformations. These transformations incorporate the conditions associated with having a determinant of the metric tensor being $1$ [equation \eqref{Eq 527}] without explicitly making this assumption. In Droste's derivation, the coordinate transformations are chosen to simplify the metric and the Einstein field equations. These transformations, while not directly stating $\sqrt{-g}=1$, equation \eqref{Eq 529}, lead to a form of the metric where the complexity that would arise from a non-unitary $\sqrt{g}$ is avoided \cite{Droste}.

Both Droste and Schwarzschild arrived at the same destination - what is now known as "the Schwarzschild metric," equation \eqref{Eq 500} - but they took somewhat different routes, reflecting their approaches to the problem of solving Einstein's field equations \eqref{Eq 30} under the assumption of spherical symmetry in a vacuum.  
Let us compare between Schwarzschild and Droste. It is important to note that this comparison is approached from a qualitative and epistemic perspective rather than a deeply technical one. This discussion explores the broader conceptual and philosophical aspects of their contributions to general relativity, shedding light on the underlying principles and ideas rather than the intricate mathematical details of their derivations.

Droste, like Schwarzschild, employed a "trick." However, it was a different trick than Schwarzschild's approach. Both Schwarzschild and Droste used such tricks in their respective derivations of the metric \eqref{Eq 500}, albeit in different ways. Both tricks are examples of mathematical ingenuity applied to the problem of finding a solution to Einstein's field equations for a spherically symmetric mass distribution.
The difference between Droste's and Schwarzschild's derivations of the metric \eqref{Eq 500} lies in their approaches to coordinate transformations and how they addressed the determinant of the metric tensor equation \eqref{Eq 527}. 

Schwarzschild started with a general line element \eqref{Eq 505} in rectangular coordinates and then transformed it into polar coordinates. He faced an issue with the transformed volume element not maintaining a determinant of $1$. Schwarzschild introduced a clever coordinate transformation equation \eqref{Eq 507}, changing the radial and angular coordinates to simplify calculations while retaining the form of the field equations \eqref{Eq 30} and the determinant condition \eqref{Eq 527}. He derived the functions $f_1, f_2, f_3$, and $f_4$ equation \eqref{Eq 504} that satisfied Einstein's conditions and the vacuum field equations \eqref{Eq 30}. Schwarzschild ensured that the product of these functions resulted in a determinant that simplified to $R^2 \sin \theta$, not $1$. He ensured physical validity by relating the integration constants $\alpha$ and $\rho$ and focused on continuity conditions \cite{Schwarzschild}.

Droste's approach involved transforming coordinates in a way that normalized certain coefficients in the metric to unity, simplifying the solution of the field equations. While Droste's method did not explicitly focus on the determinant of the metric tensor, his choice of coordinate transformations led to a simplification similar to having a unitary determinant in certain conditions. Droste's work primarily focused on obtaining a manageable form for the metric tensor components to solve Einstein's field equations under specific conditions \cite{Droste}. 

Droste's approach differs from Schwarzschild's explicit handling of the determinant condition \eqref{Eq 529}; this difference in methodology demonstrates that Droste's derivation was developed through a separate line of reasoning, independently arriving at a similar solution. Specifically, Schwarzschild's method more directly addressed the determinant [equation \eqref{Eq 527}] issue, introducing new variables \eqref{Eq 507} to transform the volume element. Droste's approach was more about simplifying the metric form through coordinate choices. 
Schwarzschild's approach explicitly accounted for the determinant to be $1$ through his transformations, which was crucial for his solution's physical validity and continuity. On the other hand, while not overtly focused on the determinant condition equation \eqref{Eq 527}, Droste's method (transformations) led to simplifications in the metric that implied a unit determinant under certain conditions.
Therefore, Schwarzschild's derivation was more explicit in handling the determinant of the metric tensor condition \eqref{Eq 529}, while Droste's approach implicitly incorporated these considerations through his choice of coordinate transformations. 

\section{Schwarzschild calculates three integrals} \label{5}

Schwarzschild assumed that space is isotropic and that the motion of a mass point in the gravitational field is confined to the equatorial plane ($\theta =  d\theta = 0$ or $x_2 =0$). He then obtained three integrals from these assumptions and his line element \eqref{Eq 500}, essential for describing the motion of the mass point in this gravitational field. Further, Schwarzschild assumed that the coefficients of the line element \eqref{Eq 500} do not depend on the time coordinate $t$ (which implies time translation symmetry) and the azimuthal angle $\phi$ (which implies rotational symmetry). The symmetries of the system (i.e., time independence and rotational symmetry) lead to conserved quantities through these equations, in accordance with Emmy Noether's theorem (for every differentiable symmetry of the action of a physical system, there is a corresponding conservation law). Thus, Schwarzschild points out that due to the homogeneity of the line element in the differentials and the symmetries in time and azimuthal angle, there are three constants of motion in the system \cite{Schwarzschild}: 

The first integral is Einstein's area law, stating that the product $R^2 \frac{d\phi}{ds}$ is constant:

\begin{equation} \label{Eq 508}
R^2 \frac{d\phi}{ds} = \text{const.} = c.
\end{equation}

\noindent The quantity $R^2 \frac{d\phi}{ds}$  (where $\phi$ is the azimuthal angle) remains constant along the geodesic, representing \emph{conservation of angular momentum} for a particle moving in this spacetime. 

\noindent Similarly, the lack of explicit time dependence in the line element implies \emph{the conservation of energy}. This gives rise to the integral involving $\frac{dt}{ds}$, which remains constant along the particle's path:

\begin{equation} \label{Eq 509}
(1 - \frac{\alpha}{R}) \frac{dt}{ds} = \text{const.} = 1.
\end{equation}

\noindent The third integral, combining the first two, comes from the normalization condition of the four-velocity (the derivative of the coordinates with respect to proper time $ds$) for timelike geodesics, with $h$ being a constant of integration. It provides a relationship between the energy, angular momentum, and the radial component of the motion: 

\begin{equation} \label{Eq 510}
(1 - \frac{\alpha}{R}) \left( \frac{dt}{ds} \right)^2 - \frac{1}{1 - \frac{\alpha}{R}} \left( \frac{dR}{ds} \right)^2 - R^2 \left( \frac{d\phi}{ds} \right)^2 = \text{const.} = h.
\end{equation}
\vspace{1mm} %1mm vertical space

The Lagrangian $\mathcal{L}$ is symmetric with respect to continuous transformation (time translation and rotation). Thus, Noether's theorem guarantees the corresponding conservation law. If the Lagrangian does not explicitly depend on time, Noether's theorem implies energy conservation equation \eqref{Eq 509}. Invariance under rotations results in the conservation of angular momentum equation \eqref{Eq 508}.  

Using the Euler-Lagrange equations, Droste derived from the line element \eqref{Eq 500} conservation laws. 
The following generally outlines Droste's derivation of Einstein's area law (the conservation of angular momentum in a spherically symmetric gravitational field) \cite{Droste}. 
\noindent For $\phi$, the Euler-Lagrange equation is expressed as:

\begin{equation} \label{Eq 545}
\frac{d}{ds}\left(\frac{\partial \mathcal{L}}{\partial \dot{\phi}}\right) - \frac{\partial \mathcal{L}}{\partial \phi} = 0,
\end{equation}

\noindent where $\dot{\phi}$ is the derivative of $\phi$ with respect to the proper time $s$ and $\mathcal{L}$ is the Lagrangian of the system.

\noindent In the Schwarzschild metric, the Lagrangian $\mathcal{L}$  does not explicitly depend on the azimuthal angle $\phi$, which means that:

\begin{equation} \label{Eq 546}
\frac{\partial \mathcal{L}}{\partial \phi} = 0.    
\end{equation}

\noindent Given equation \eqref{Eq 546}, the Euler-Lagrange equation \eqref{Eq 545} simplifies to: 

\begin{equation}
\frac{d}{ds}\left(\frac{\partial \mathcal{L}}{\partial \dot{\phi}}\right) = 0.
\end{equation}

\noindent This implies that $\frac{\partial \mathcal{L}}{\partial \dot{\phi}}$ is a constant of motion. 
Given the form of the Schwarzschild Lagrangian, $\frac{\partial \mathcal{L}}{\partial \dot{\phi}}$ is proportional to:

\begin{equation}
R^2 \Dot{\phi} = R^2\frac{d \phi}{ds}.     
\end{equation}

\noindent Thus, $R^2\frac{d \phi}{ds}$ is a constant and we obtain equation \eqref{Eq 508}.

Schwarzschild published his solution in 1916. This was two years before Emmy Noether published her groundbreaking paper in 1918. Though Schwarzschild's work predated Noether's theorem, the concepts of symmetries and conservation laws were already central to physics. However, Noether's work provided a more formal and general mathematical framework for understanding these concepts \cite{Noether}. 

\section{Schwarzschild deduces Einstein's formula for the orbit of Mercury} \label{6}

Schwarzschild showed that the integrals \eqref{Eq 508}, \eqref{Eq 509}, \eqref{Eq 510} lead to the correct prediction of Mercury's perihelion advance, confirming the validity of his solution \eqref{Eq 500}. Schwarzschild manipulated the integrals \eqref{Eq 508}, \eqref{Eq 509}, and \eqref{Eq 510} and obtained Mercury's orbit equation in the Sun's gravitational field. 
Let us derive Einstein's equation for the planet's orbit from the three integrals.  
First, we isolate $\frac{dt}{ds}$ from equation \eqref{Eq 509}:

\begin{equation} \label{Eq 519}
\frac{dt}{ds} = \frac{1}{1-\frac{\alpha}{R}}.  
\end{equation}

\noindent Then, we isolate $\frac{d \phi}{ds}$ from the area law, equation \eqref{Eq 508}: 

\begin{equation} \label{Eq 520}
\frac{d \phi}{ds} = \frac{c}{R^2}    
\end{equation}

\noindent This expression involves squaring the fraction $\frac{c}{R^2}$: $\frac{c^2}{R^4}$. 

\noindent We then plug equations \eqref{Eq 519} and \eqref{Eq 520} into equation \eqref{Eq 510}. 

\noindent Schwarzschild wanted to express everything in terms of $\frac{dR}{d\phi}$. 

\noindent We use the chain rule:

\begin{equation} \label{Eq 534}
\frac{dR}{ds} = \frac{dR}{d\phi} \frac{d\phi}{ds}.    
\end{equation}

\noindent Substituting equation \eqref{Eq 520} into the expression \eqref{Eq 534} we get:

\begin{equation} \label{Eq 535}
\frac{dR}{ds} = \frac{dR}{d\phi} \frac{c}{R^2}.    
\end{equation}

\noindent Substituting equations \eqref{Eq 519} and \eqref{Eq 535} back into the equation \eqref{Eq 510}, we can solve for $\frac{dR}{d\phi}$. We get:

\begin{equation} 
\frac{1}{(1 - \frac{\alpha}{R})} - \frac{1}{1 - \frac{\alpha}{R}} \left( \frac{dR}{d \phi} \right)^2\frac{c^2}{R^4}-R^2\left(\frac{d \phi}{ds}\right)^2 = h
\end{equation}

\noindent Since $\frac{1}{(1 - \frac{\alpha}{R})}$ is the reciprocal of $(1 - \frac{\alpha}{R})$, multiplying these terms together simplifies to $1$. Thus multiplying by $(1 - \frac{\alpha}{R})$, we get the following:

\begin{equation} 
 - \left( \frac{dR}{d \phi} \right)^2\frac{c^2}{R^4} -R^2\left({1 - \frac{\alpha}{R}}\right)\frac{c^2}{R^4} = h\left({1 - \frac{\alpha}{R}}\right)-1.
\end{equation}

After substitution and rearrangement, 
Schwarzschild arrived at an orbit equation similar to that of Einstein for the perihelion precession of Mercury \cite{Schwarzschild}:

\begin{equation} \label{Eq 547}
\left( \frac{dR}{d\phi} \right)^2 + R^2 \left(1 - \frac{\alpha}{R}\right) = \frac{R^4}{c^2} \left[1 - h \left(1 - \frac{\alpha}{R}\right)\right].
\end{equation}

Schwarzschild then wrote the orbit equation in terms of a new variable $x = \frac{1}{r} \equiv \frac{1}{R}$ expressing his equation in accordance with Einstein's notation \cite{Schwarzschild}:

\begin{equation}  \label{Eq 548}
\begin{aligned}
\left( \frac{dx}{d\phi} \right)^2 &= \frac{1-h}{c^2} + \frac{h\alpha}{c^2}x - x^2 + \alpha x^3. 
\end{aligned}  
\end{equation}

\noindent This equation captures the key features of Einstein's orbit equation, including the precession of Mercury's perihelion.

Let us derive equation \eqref{Eq 548} from equation \eqref{Eq 547}. We start with the left-hand side of the orbit equation \eqref{Eq 547} and substitute $R=\frac{1}{x}$:

\begin{equation} \label{Eq 551}
R^2 \left(1 - \frac{\alpha}{R}\right) = \frac{1}{x^2} \left(1 - \alpha x\right).    
\end{equation}

\noindent The transformation of the derivative part of the orbit equation is:

\begin{align}
dR &= -\frac{dx}{x^2} \
\frac{dR}{d\phi} = -\frac{1}{x^2} \frac{dx}{d\phi}. \quad \text{Thus:}
\end{align}

\begin{align} \label{Eq 549}
\left( \frac{dR}{d\phi} \right)^2 = \left( -\frac{1}{x^2} \frac{dx}{d\phi} \right)^2 \
&= \frac{1}{x^4} \left( \frac{dx}{d\phi} \right)^2.
\end{align}

\noindent For the right-hand side, we again substitute $R=\frac{1}{x}$ and simplify:

\begin{align} \label{eq 550}
\frac{R^4}{c^2} \left[1 - h \left(1 - \frac{\alpha}{R}\right)\right] &= \frac{1}{c^2 x^4} \left[1 - h \left(1 - \alpha x\right)\right].
\end{align}

\noindent Now, we substitute equations \eqref{Eq 549}, \eqref{Eq 551} and \eqref{eq 550} into the original orbit equation \eqref{Eq 547} and obtain:

\begin{align}
\frac{1}{x^4} \left( \frac{dx}{d\phi} \right)^2 + \frac{1}{x^2} \left(1 - \alpha x\right) &= \frac{1}{c^2 x^4} \left[1 - h \left(1 - \alpha x\right)\right].
\end{align}

\noindent Next, we multiply through by $x^4$ to eliminate the denominators:

\begin{align} \label{Eq 591}
\left( \frac{dx}{d\phi} \right)^2 + x^2 - \alpha x^3 &= \frac{1}{c^2} - \frac{h}{c^2} + \frac{h\alpha x}{c^2} = \frac{1-h}{c^2} + \frac{h\alpha}{c^2}x - x^2 + \alpha x^3.
 \end{align}

\noindent Finally, we rearrange and obtain equation \eqref{Eq 548}.
Equation \eqref{Eq 591} matches Einstein’s formulation from his Mercury perihelion paper \cite{Einstein15}:

\begin{equation} \label{Eq 68-1}
\left( \frac{dx}{d\phi} \right)^2 = \frac{2A}{B^2} + \frac{\alpha}{B^2} x - x^2 - \alpha x^3, 
\end{equation}

\begin{equation*}
\text{only if:    } \ \frac{2A}{B^2} = \frac{1-h}{c^2} \             
\text{and } B^2 = \frac{c^2}{h},      
\end{equation*}
\vspace{1mm} %1mm vertical space

\noindent verifying the anomaly in Mercury’s orbit that had puzzled astronomers for decades \cite{Schwarzschild}.
Setting $x=\frac{1}{r}$ and $A = E$ equation \eqref{Eq 68-1} can be written as follows:

\begin{equation} \label{Eq 68}
\left(\frac{d}{d\phi} \left(\frac{1}{r}\right)\right)^2 = \frac{2E}{B^2} + \frac{\alpha}{rB^2} - \frac{1}{r^2} + \frac{\alpha}{r^3}.  
\end{equation}
\vspace{1mm} %1mm vertical space

Schwarzschild notes that Einstein's approximation for the orbit of a planet (like Mercury) aligns with the exact solution \eqref{Eq 500} when substituting for $r$ in the equation \eqref{Eq 68} the relationship equation \eqref{Eq 536} \cite{Schwarzschild}: 

\begin{equation} \label{Eq 536a}
R = \left(r^3 + \alpha^3\right)^{\frac{1}{3}} \approx r \left( 1 + \frac{\alpha^3}{r^3}\right),   
\end{equation}

\noindent valid when $\frac{\alpha^3}{r^3}$ is much less than $1$. 

Schwarzschild notes that in the case of an approximate solution like Einstein's, the term $\frac{\alpha^3}{r^3}$ is extremely small. Recall that the term $\alpha$ is defined by equation \eqref{Eq 73}. Schwarzschild points out that $\frac{\alpha}{r}$ is nearly equal to twice the square of the velocity of the planet (when the speed of light = $1$), which is a small fraction. Due to the smallness of $\frac{\alpha^3}{r^3}$, the expression $\left( 1 + \frac{\alpha^3}{r^3}\right)$ on the right-hand side of equation \eqref{Eq 536a} is very close to 1. Therefore, for practical purposes, $R$ is almost identical to $r$ as the difference between $r$ and $R$ is extremely small. In such cases, $R \approx r$ is extremely accurate. Schwarzschild affirms that Einstein's approximation, which treats $R$ as equivalent to $r$ in the approximate methodology, is more than sufficient for the accuracy required in practical applications, like calculating the orbit of Mercury. The difference between $\left( 1 + \frac{\alpha^3}{r^3}\right)$ and $1$ is so minute (of the order $10^{-12}$) that it does not significantly impact the calculations of Mercury's orbit to a high degree of precision.

\section{Schwarzschild employs the same trick for an internal solution} \label{7}

After reading Einstein's November 25 paper, which extended his November 18 field equations, Schwarzschild approached the singularity problem in his line element. Acquainting himself with Einstein's energy tensor, Schwarzschild extended his calculation from the gravitational field of a mass point to the gravitational field of an incompressible homogeneous and isotropic fluid sphere. Utilizing Einstein's revised field equations from his November 25 paper, Schwarzschild derived an exact solution for the interior of the incompressible fluid sphere. Schwarzschild's further research demonstrated that his line element \eqref{Eq 500} applied to the region outside an incompressible fluid sphere.

Schwarzschild communicated his new findings to Einstein in a letter dated February 6, 1916, attaching his paper titled "On the Gravitational Field of a Sphere of Incompressible Fluid, According to Einstein's Theory"\cite{Einstein10} \cite{CPAE8}, Doc. 188. 
Einstein's reply to Schwarzschild's letter came on February 19, and he then presented Schwarzschild's paper to the Prussian Academy on February 24, 1916.

In his further research, Schwarzschild began with Einstein's field equations valid only in unimodular coordinates \eqref{Eq 529} \cite{Einstein14}:

\begin{equation} \label{Eq 20}
G_{\mu\nu} = -\kappa \left( T_{\mu\nu} - \frac{1}{2} g_{\mu\nu} T \right), \sqrt{-g}=1.    
\end{equation}

\noindent $G_{\mu\nu}$ is the Einstein tensor, $T_{\mu\nu}$ is the energy-momentum tensor, and $g_{\mu\nu}$ is the metric tensor. 

These equations \eqref{Eq 20} simplify in regions without matter (outside the fluid sphere), i.e., the components of $G_{\mu\nu}$ vanish for areas without matter [see equations \eqref{Eq 30}]. Inside the fluid sphere, the components of $G_{\mu\nu}$ are defined using the mixed stress-energy tensor of an incompressible fluid, with the pressure $p$ and constant density $\rho_0$ as: 

\begin{equation} \label{Eq 538}
T_1^1 = T_2^2 = T_3^3 = -p, \quad T_4^4 = \rho_0,
\end{equation}
\vspace{1mm} %1mm vertical space 

\noindent and all other components of $T_\mu^\nu$ are zero. 

In his discussion about equation \eqref{Eq 538}, Schwarzschild mentions "Einstein's 'mixed energy tensor'" while referencing Einstein's 1914 \emph{Entwurf} review paper, "The Formal Foundation of the General Theory of Relativity." Einstein devised his stress-energy tensor of a perfect fluid, defined in terms of pressure $p$ and constant density $\rho_0$ in the following manner \cite{Einstein11}:

\begin{equation} \label{Eq 550}
\begin{pmatrix}
-p & 0 & 0 & 0 \\
0 & -p & 0 & 0 \\
0 & 0 & -p & 0 \\
0 & 0 & 0 & \rho_0(1 + P)
\end{pmatrix}.
\end{equation}

\noindent In this equation, $p$ represents the pressure in the stress-energy tensor, $P$ is a derived quantity from $p$: an integral that is connected with the energy content of the system, i.e., $\rho_0(1 + P)$ represents the energy density. 

Schwarzschild describes the structure of the tensor $T_\mu^\nu$, where one index is covariant (subscript) and the other is contravariant (superscript). This specific arrangement of indices in the tensor illustrates the "mixed" nature of the energy tensor per Einstein's formulation in his 1914 paper. However, in his 1914 \emph{Entwurf} review paper, Einstein referred to his stress-energy tensor $T^\nu_\sigma$ as the stress-energy tensor density. A tensor density incorporates the factor $\sqrt{-g}$. The $\sqrt{-g}$ allows the tensor density to transform differently under coordinate transformations than a regular tensor. However, in the case of the tensor equation \eqref{Eq 550}, we would not include the $\sqrt{-g}$. Hence, in the specific case of the matrix representing a perfect fluid \eqref{Eq 550}, the distinction between the "mixed" stress-energy tensor and the tensor "density" is not immediately relevant.

To achieve equilibrium, the fluid sphere must satisfy Einstein's equation for energy-momentum balance for matter in a gravitational field \cite{Einstein1}:

\begin{equation} \label{eq 212}
\sum_{\alpha}\frac{\partial T_\sigma^\alpha}{\partial x_\alpha} = -\sum_{\alpha\beta}\Gamma^\alpha_{\sigma\beta} T_\alpha^\beta.
\end{equation}

\noindent On the right-hand side, we have the Christoffel symbols, the components of the gravitational field $\Gamma^\alpha_{\sigma\beta}$, equation \eqref{Eq 278}. The equilibrium condition \eqref{eq 212} relates the divergence of the stress-energy tensor $T_\sigma^\alpha$ to the Christoffel symbols and the stress-energy tensor $\Gamma^\alpha_{\sigma\beta} T_\alpha^\beta$. 

Recall that Schwarzschild transformed the usual spherical coordinates into new coordinates to simplify the mathematical treatment; see equations \eqref{Eq 507}, section \ref{3}. The transformation  \eqref{Eq 507} is chosen to facilitate the application of Einstein's field equations in a spherically symmetric scenario. With these transformed coordinates, Schwarzschild writes the line element \eqref{Eq 530}($ds^2$) in a form adapted to Einstein's four conditions and the determinant condition equation \eqref{Eq 527}, where $f_1, f_2$, and $f_4$ are functions of $x_1$. 
The line element directly leads to the components of the metric tensor $g_{\mu \nu}$ \eqref{Eq 517}. These components and equations \eqref{Eq 504a}, \eqref{Eq 504b}, and \eqref{Eq 574} reflect the curvature of spacetime outside the fluid sphere.   

Schwarzschild uses the line element \eqref{Eq 530} and the associated metric tensor \eqref{Eq 517} to apply Einstein's field equations \eqref{Eq 20} specifically to the interior of the incompressible fluid sphere. This involves solving the field equations using the derived metric components, considering the properties of the perfect fluid (i.e., considering $p$ and $\rho_0$) and equation \eqref{Eq 538} \cite{Einstein10}:

\begin{equation}
T_{11} = T_{1}^{2} (-pg_{11}), \ T_{22} = T_{2}^{2} (-pg_{22}), \ T_{33} = T_{3}^{2} (-pg_{33}).\ T_{44} = T_{4}^{2} (\rho_0 g_{44}),
\end{equation}

\noindent where $g_{11}, g_{22}, g_{33}$ and $ g_{44}$ are represented by equations \eqref{Eq 517}. And:

\begin{equation}
G_{11} = \frac{\kappa f_2}{2}(p - \rho_{0}), \ G_{22} = \frac{\kappa f_2}{2}\frac{1}{1 - x_{2}^{2}}(p - \rho_{0}).
\end{equation}

\begin{equation*}
G_{33} = \frac{\kappa f_2}{2}(1-x^2_2)(p - \rho_{0}), \ G_{44} = -\frac{\kappa f_4}{2}(\rho_{0} + 3p).    
\end{equation*}
\vspace{1mm} %1mm vertical space

Schwarzschild's derivation aims to find a solution for the sphere's interior that matches smoothly with the exterior Schwarzschild's solution \eqref{Eq 500} at the sphere's boundary. This matching guarantees a consistent overall description of spacetime around and inside the sphere.    

Schwarzschild considers the Christoffel symbols in terms of the functions $f$ from his previous paper \cite{Schwarzschild} equations \eqref{Eq 572a}, \eqref{Eq 572b}, and \eqref{Eq 572c}, which involve only derivatives of the functions $f_1, f_2, f_3, f_4$ with respect to $x_1$. Recall that in \cite{Schwarzschild}, Schwarzschild set $x_2=0$ at the equator as a simplifying assumption streamlining the field equations because the equations involve rotational symmetry around the origin. In his second paper, again at the equator, he sets $x_2$ = 0, thus:

\begin{equation} \label{Eq 518c}
-pg_{22} = -pf_{2} \text{ (at } x_2 = 0 \text{)}, 
\end{equation}

\noindent and the simplified field equations \cite{Einstein10}:

\begin{equation} \label{Eq 518}
- \frac{1}{2} \frac{\partial}{\partial x_{1}} \left( \frac{1}{f_{1}} \frac{\partial f_{1}}{\partial x_{1}} \right) + \frac{1}{4} \frac{1}{f_{1}^2} \left( \frac{\partial f_{1}}{\partial x_{1}} \right)^{2} + \frac{1}{2} \frac{1}{f_{2}^2} \left( \frac{\partial f_{2}}{\partial x_{1}} \right)^{2} + \frac{1}{4} \frac{1}{f_{4}^2} \left( \frac{\partial f_{4}}{\partial x_{1}} \right)^{2}
\end{equation}

\begin{equation*}
= - \frac{\kappa}{2}f_1(\rho_{0} - p).    
\end{equation*}

\begin{equation} \label{Eq 518a}
\frac{1}{2} \frac{\partial}{\partial x_{1}} \left( \frac{1}{f_{1}} \frac{\partial f_{2}}{\partial x_{1}} \right) - 1 - \frac{1}{2} \frac{1}{f_{1} f_{2}} \left( \frac{\partial f_{2}}{\partial x_{1}} \right)^{2} = - \frac{\kappa}{2}f_2(\rho_{0} - p).
\end{equation}

\begin{equation} \label{Eq 518b}
- \frac{1}{2} \frac{\partial}{\partial x_{1}} \left( \frac{1}{f_{1}} \frac{\partial f_{4}}{\partial x_{1}} \right) + \frac{1}{2} \frac{1}{f_{1} f_{4}} \left( \frac{\partial f_{4}}{\partial x_{1}} \right)^{2} = - \frac{\kappa}{2}f_4 (\rho_{0} + 3p).
\end{equation}
\vspace{1mm} %1mm vertical space 

\noindent The field equations relate the partial derivatives of the functions $f_1, f_2$, and $f_4$ to the pressure $p$ and the constant density $\rho_0$. 

Similar to the derivation of the external line element \eqref{Eq 500}, the determinant condition \eqref{Eq 574} plays a significant role in Schwarzschild's derivation of the line element for the interior of an incompressible fluid sphere; it is useful for simplifications and substitutions. Recall that Schwarzschild derived an additional condition \eqref{Eq 575} (also called the determinant condition), see section \ref{3}, which relates the derivatives of the functions $f_1, f_2$, and $f_4$ to each other, providing a direct way to connect changes in these functions. 

Now, Schwarzschild wanted to rewrite the equilibrium condition \eqref{eq 212} in terms of the properties of the perfect fluid sphere $p$ and $\rho_0$. For this purpose, he followed the following steps:

1) He started with the equilibrium condition \eqref{eq 212} and expressed the divergence of the stress-energy tensor $\frac{\partial T_\sigma^\alpha}{\partial x_\alpha}$ in terms of the metric functions $f_1, f_2$, and $f_4$ and their derivatives. 
These metric functions correspond to the components of the metric tensor: $-pg_{11}, -pg_{22}, -pg_{33}$ and $ \rho_0 g_{44}$ [see equation \eqref{Eq 517}]. Remember that the derivatives are with respect to the coordinate $x_1$.
This process involves the Christoffel symbols equations \eqref{Eq 572a}, \eqref{Eq 572b}, and \eqref{Eq 572c}. In other words, Schwarzschild used the equilibrium condition \eqref{eq 212} to connect the pressure $p$ and constant density $\rho_0$
with the geometry of spacetime embodied in the Christoffel symbols.

2) Schwarzschild considered a simplified scenario where $x_2 =0$ and applied the simplified field equations \eqref{Eq 518}, \eqref{Eq 518a}, \eqref{Eq 518b} to replace expressions involving $p$ and $\rho_0$ with expressions involving the derivatives of the metric functions $f_1, f_2$, and $f_4$. 

3) Schwarzschild then employs the determinant condition \eqref{Eq 574} to simplify expressions and substitute one metric function in terms of others. He uses the additional (determinant) condition equation \eqref{Eq 575} to simplify and eliminate redundancies. Recall that this equation relates the derivatives of the functions $f_1, f_2$, and $f_4$ to each other and connects changes in these functions, allowing for further simplifications in the derivation.

Combining the components 1), 2), and 3), we arrive at a new form of the equilibrium condition \cite{Einstein10}: 

\begin{equation} \label{Eq 518d}
-\frac{\partial p}{\partial x_1} = -\frac{p}{2} \left[ \frac{1}{f_1} \frac{\partial f_1}{\partial x_1} + \frac{2}{f_2} \frac{\partial f_2}{\partial x_1}\right] + \frac{\rho_0}{2} \frac{1}{f_4} \frac{\partial f_4}{x_1}.    
\end{equation}

\noindent This equilibrium condition expresses $p$ and $\rho_0$ in terms of the derivatives of the $f$ functions.

\noindent With the determinant condition \eqref{Eq 574}, the equilibrium condition \eqref{Eq 518d} becomes:

\begin{equation}
-\frac{\partial p}{\partial x_1} = \frac{\rho_0 +p}{2} \frac{1}{f_4}\frac{\partial f_4}{\partial x_1}.    
\end{equation}

\noindent This equation relates the pressure gradient to the constant density $\rho_0$, the pressure $p$, and the derivative of the metric function $f_4$. 

\noindent Now, integrating both sides with respect to $x_1$ gives:

\begin{equation} \label{Eq 586}
(\rho_0 +p)\sqrt{f_4} =const. = \gamma, \text{ where } \gamma \text{ is a constat.}   
\end{equation}

Schwarzschild then multiplies each of the field equations \eqref{Eq 518}, \eqref{Eq 518a}, \eqref{Eq 518b} by specific factors and simplifies them: he multiplies equation \eqref{Eq 518} by $-2$ and this multiplication cancels the negative sign and the factor of $\frac{1}{2}$ in the first term, simplifying the equation. Schwarzschild then multiplies equation \eqref{Eq 518a} by $2\frac{f_1}{f_2}$. This multiplication is intended to simplify the terms involving $f_2$, particularly the first term on the left-hand side. He multiplies the third equation \eqref{Eq 518b} by $-2\frac{f_1}{f_4}$. Similar to the second field equation, this multiplication simplifies terms involving $f_4$. The resulting transformed equations are (where for simplicity $x_1 \equiv x$) \cite{Einstein10}:

\begin{equation} \label{Eq 518e}
\frac{\partial}{\partial x} \left( \frac{1}{f_{1}} \frac{\partial f_{1}}{\partial x} \right) = \frac{1}{2f^2_{1}} \left( \frac{\partial f_{1}}{\partial x} \right)^{2} + \frac{1}{f_{2}^{2}} \left( \frac{\partial f_{2}}{\partial x} \right)^{2} + \frac{1}{2f_{4}^{2}} \left( \frac{\partial f_{4}}{\partial x} \right)^{2} + \kappa f_{1}(\rho_{0} - p).
\end{equation}

\begin{equation} \label{Eq 518f}
\frac{\partial}{\partial x} \left( \frac{1}{f_{2}} \frac{\partial f_{2}}{\partial x} \right) = 2\frac{f_{1}}{f_{2}} + \frac{1}{f_{1} f_{2}} \frac{\partial f_{1}}{\partial x} \frac{\partial f_{2}}{\partial x} - \kappa f_{1}(\rho_{0} - p).
\end{equation}

\begin{equation} \label{Eq 518g}
\frac{\partial}{\partial x} \left( \frac{1}{f_{4}} \frac{\partial f_{4}}{\partial x} \right) = \frac{1}{f_{1} f_{4}} \frac{\partial f_{1}}{\partial x} \frac{\partial f_{4}}{\partial x} + \kappa f_{1}(\rho_{0} + 3p).
\end{equation}
\vspace{1mm} %1mm vertical space 

The next step was to combine the transformed field equations \eqref{Eq 518e}, \eqref{Eq 518f}, \eqref{Eq 518g} in such a way that they simplify further, using the determinant condition \eqref{Eq 574} to facilitate this process. 

First, Schwarzschild combined equation \eqref{Eq 518e}, twice equation \eqref{Eq 518f}, and equation \eqref{Eq 518g}. The determinant condition \eqref{Eq 574} 
plays a role in eliminating certain terms and simplifying the equations further. Subsequently, again, Schwarzschild combined equation \eqref{Eq 518e} and equation \eqref{Eq 518g} similarly. The process was similar to the first combination, where he aligned terms involving the derivatives and used the determinant condition \eqref{Eq 574} to simplify. After combining and simplifying, he arrived at two equations \cite{Einstein10}: 

\begin{equation} \label{Eq 587a}
0 = 4\frac{f_1}{f_2} - \frac{1}{f_2^2} \left( \frac{\partial f_2}{\partial x} \right)^2 - 2\frac{1}{f_2 f_4} \frac{\partial f_2}{\partial x} \frac{\partial f_4}{\partial x} + 4\kappa f_1 p
\end{equation}

\begin{equation} \label{Eq 587b}
0 = 2 \frac{\partial}{\partial x} \left( \frac{1}{f_2} \frac{\partial f_2}{\partial x} \right) + 3\frac{1}{f_2^2} \left( \frac{\partial f_2}{\partial x} \right)^2 + 2\kappa f_1(\rho_0 + p)
\end{equation}
\vspace{1mm} %1mm vertical space
 
\section{Schwarzschild's again reverts to spherical coordinates}

I will streamline the derivation because I want to highlight only the key steps relevant to the subject matter of this paper. So, the field equations \eqref{Eq 587a}, \eqref{Eq 587b}, and the equilibrium condition \eqref{Eq 586} finally lead to the following values for the functions \cite{Einstein10}:

\begin{equation} \label{Eq 592}
f_2 = \frac{3}{\kappa \rho_0} sin^2 \sigma, f_4 = \left(\frac{3 \cos \sigma_{\text{ext}} - \cos \sigma}{2}\right)^2, \text{ valid with the condition \eqref{Eq 574}}.     
\end{equation}
\vspace{1mm} %1mm vertical space 

Schwarzschild combined $f_4$ [equation \eqref{Eq 592}], and the equilibrium condition, equation \eqref{Eq 586}, and obtained an equation that relates the pressure and density inside the static, spherically symmetric body (i.e., a non-rotating star).

\noindent From the equilibrium equation, we have: $\rho_0+p=\frac{\gamma}{\sqrt{f_4}}$. Substituting $f_4$ from equation \eqref{Eq 592} we get:

\begin{equation} \label{Eq 605}
\rho_0+p=\frac{\gamma}{\frac{3 \cos \sigma_{\text{ext}} - \cos \sigma}{2}}.    
\end{equation}

\noindent Let us consider the boundary condition where $\sigma =\sigma_{\text{ext}}$. At this boundary, the equilibrium condition, equation \eqref{Eq 586}, must hold. Plugging $\sigma =\sigma_{\text{ext}}$ into $f_4$ [equation \eqref{Eq 592}] we get: 

\begin{equation} \label{Eq 592a}
f_4 = \left(\frac{3 \cos \sigma_{\text{ext}} - \cos \sigma_{\text{ext}}}{2}\right)^2 = \cos \sigma^2_{\text{ext}}.     
\end{equation}

\noindent Therefore, at the boundary, the equilibrium condition, equation \eqref{Eq 586} becomes: 

\begin{equation} \label{Eq 607}
(\rho_0+p)\cos \sigma_{\text{ext}}= \gamma.    
\end{equation}

\noindent Assuming $p=0$ at the boundary, we get from equation \eqref{Eq 607}:

\begin{equation} \label{Eq 608}
\rho_0 \cos \sigma_{\text{ext}}= \gamma.    
\end{equation}

\noindent Substituting equation \eqref{Eq 608} back into equation \eqref{Eq 605}, we get \cite{Einstein10}:

\begin{equation} \label{Eq 601}
\rho_0+p=\rho_0\frac{2\cos \sigma_{\text{ext}}}{3 \cos \sigma_{\text{ext}} - \cos \sigma}.   
\end{equation}

\noindent $\rho_0$, the constant density, and $p$, the pressure, are measured within the spherically symmetric body. The equation describes how the pressure $p$ changes with the radial distance inside a star.

The specific forms of $f_2$ and $f_4$ in equation \eqref{Eq 592} are designed to describe the spacetime geometry inside the fluid sphere.
Schwarzschild defines the functions $f_2$ and $f_4$ in terms of the spherical variable $\sigma$ and the constant parameters $\kappa$ (Einstein's gravitational constant) and $\rho_0$. $\sigma= 3x - \rho$ is valid outside the sphere, and as said above, $\sigma = \sigma_{\text{ext}}$ at the sphere's surface. The functions $f_2$ and $f_4$ are metric tensor components in the interior line element \eqref{Eq 514}. 

After establishing the solution in the coordinates \eqref{Eq 507}, Schwarzschild then transforms back to standard spherical coordinates  $\sigma, \theta, \phi$. While spherical coordinates do not satisfy the determinant condition \eqref{Eq 574}, the derived functions $f_2$ and $f_4$ [\eqref{Eq 592}] are still applicable in spherical coordinates. Schwarzschild's skillful application of the determinant condition \eqref{Eq 574} and Cartesian coordinates and subsequent transformation back to spherical coordinates demonstrate his ability to navigate the complexities of Einstein's 1915 formulation.

Using spherical variables $\sigma, \theta, \phi$ instead of $x_1, x_2, x_3$, Schwarzschild writes the line element for the region inside the sphere, showing it to be free of singularities \cite{Einstein10}: 

\begin{equation} \label{Eq 514}
ds^2 = \left(\frac{3 \cos \sigma_{\text{ext}} - \cos \sigma}{2}\right)^2 dt^2 - \frac{3}{\kappa \rho_0} d \sigma^2 + \frac{3}{\kappa \rho_0}{{\sin^2}\sigma} (d\theta^2 + \sin^2\theta d\phi^2).
\end{equation}

\noindent where the first term $\left(\frac{3 \cos \sigma_{\text{ext}} - \cos \sigma}{2}\right)^2 dt^2$, is related to the metric tensor's time-time component, which describes how gravity influences time within a spherical, non-rotating body of uniform density, like a star. The metric's second term represents the metric tensor's spatial components. It is related to the space geometry inside a spherically symmetric, non-rotating body of uniform density (i.e., a star).
The terms $\sigma, \theta, \phi$ involving these coordinates do not introduce singularities as long as the density $\rho_0$, and the fluid characteristics are well-defined and non-zero. 
The Schwarzschild interior solution \eqref{Eq 514} is regular everywhere inside the incompressible sphere (for $0\le \sigma \le \sigma_{\text{ext}}$), including at the center (where $\sigma = 0$). The metric behaves regularly at the center, and there's no physical singularity. 

\section{Schwarzschild's collapsing sphere} \label{8}

Schwarzschild analyzed the pressures within and outside a collapsing sphere.
He defined two parameters \cite{Einstein10}: 

\textsc{The first parameter $\alpha$:}

\begin{equation} \label{Eq 516}
\alpha = \sqrt{\frac{3}{\kappa \rho_0}} \sin^3 \sigma_{\text{ext}},
\end{equation}
\vspace{1mm} %1mm vertical space 

\noindent This is a key parameter defined in terms of the fluid's density $\rho_0$ and the spherical variable $\sigma_{\text{ext}}$ in equation \eqref{Eq 514}. 

\noindent Next, recall that at the boundary $\sigma = \sigma_{\text{ext}}$. Schwarzschild therefore defined \cite{Einstein10}: 

\begin{equation} \label{Eq 595}
R_{\text{ext}} = \sqrt{\frac{3}{\kappa \rho_0}} \sin \sigma_{\text{ext}},  \ \text{ thus:}  
\end{equation}
\vspace{1mm} %1mm vertical space 

\begin{equation} \label{Eq 596}
\frac{3}{\kappa \rho_0} = \frac{R_{\text{ext}}^2}{{\sin^2}\sigma_{\text{ext}}}, \ \text{ and:} 
\end{equation}

\begin{equation} \label{Eq 597}
\frac{\alpha}{R_{\text{ext}}}={{\sin^2}\sigma_{\text{ext}}}.    
\end{equation}

\noindent We can substitute ${{\sin^2}\sigma_{\text{ext}}}$ from equation \eqref{Eq 597} into equation \eqref{Eq 596}. We get:

\begin{equation} \label{Eq 598}
\frac{3}{\kappa \rho_0} = \frac{{R_{\text{ext}}}^3}{\alpha}.    
\end{equation}

Now if we combine equations \eqref{Eq 595} and \eqref{Eq 516}, we get at the boundary, equation \eqref{Eq 598}:

\begin{equation}
\alpha = \frac{\kappa \rho_0}{3}{R_{\text{ext}}}^3.     
\end{equation}

\textsc{The second parameter $\rho$:}

\noindent $\rho$ is given by: 

\begin{equation}
\rho = A^{-3/2} \left[\frac{3}{2} \sin^3 \sigma_{\text{ext}} - \frac{9}{4} \cos \sigma_{\text{ext}} B\right],
\end{equation}
\vspace{1mm} %1mm vertical space 

\noindent relating it to the sphere's density and radius, where:

\begin{equation*}
A = \frac{\kappa \rho_0}{3} \quad \text{ and:  }  B = \sigma_{\text{ext}} - \frac{1}{2} \sin 2\sigma_{\text{ext}}.  
\end{equation*}

When describing the internal properties of the sphere, Schwarzschild stressed that the characteristics of spherical geometry dominate. The curvature radius of this spherical space is established as \cite{Einstein10}:   

\begin{equation}
R_{\text{int}} = \sqrt{\frac{3}{\kappa \rho_0}} \sigma_{\text{ext}}.    
\end{equation}
\vspace{1mm} %1mm vertical space 

\noindent $R_{\text{int}}$ denotes the sphere's radius as measured inside from the sphere's center to the boundary.

Schwarzschild considered an object that can fall freely from rest at an infinite distance to the surface of the spherical sphere. The object starts from rest at an infinite distance. As the object falls towards the sphere, it follows a geodesic path, and its velocity increases due to the gravitational attraction of the sphere. Upon reaching the sphere's surface, the object attains its maximum velocity. This final velocity is calculated using the metric [equation \eqref{Eq 514}]. Schwarzschild calls this velocity the "naturally measured" fall velocity \cite{Einstein10}:

\begin{equation} \label{Eq 602}
v_{\text{ext}} = \sqrt{\frac{\alpha}{R_{\text{ext}}}}.
\end{equation}

\noindent Plugging equation \eqref{Eq 597} into this equation, we get:

\begin{equation} \label{Eq 603}
v_{\text{ext}} = \sin{\sigma_{\text{ext}}}.
\end{equation}

Schwarzschild observed that as the sphere collapses at a constant mass but increasing density, it transitions to a smaller radius than before and emits radiant energy. Radiation that is emitted from the surface of the massive sphere is redshifted according to the gravitational redshift formula:

\begin{equation} 
z={\frac {1}{\cos \sigma}-1}.    
\end{equation}

The velocity of light within the sphere is given by \cite{Einstein10}:

\begin{equation}
v_{\text{int}} =\frac{2}{3 \cos \sigma_{\text{ext}} - \cos \sigma}.  
\end{equation}

\noindent This equation indicates that the velocity of light varies within the sphere, depending on the position (as represented by $\sigma_{\text{ext}}$ and $\cos \sigma$). 
\noindent $\frac{1}\cos\sigma_{\text{ext}}$ represents the speed of light at the sphere's surface. As we move toward the center of the sphere, the speed of light increases, reaching a maximum value of $\frac{2}{3 \cos \sigma_{\text{ext}} - 1}$ at the center.  

At the sphere's center, where $\cos \sigma_{\text{ext}} = \frac{1}{3}$, \text{and} $\sigma =0$, the velocity of light reaches infinity;
the velocity of fall $v_{\text{ext}}$ [equation \eqref{Eq 602}] approaches approximately $0.94c$, where $c$ is the speed of light, signifying a density limit beyond which the sphere collapses. Equation \eqref{Eq 601} gives:  

\begin{equation} \label{Eq 601a}
\rho_0+p=\frac{2}{3}\rho_0\frac{1}{1 - 1} = 0.   
\end{equation}

\noindent The pressure at the center is infinite, and beyond this critical density point, an incompressible fluid sphere cannot sustain its structure and collapses. Schwarzschild thought that he had thus set a boundary beyond which an incompressible fluid sphere ceases to exist. For $\cos \sigma_{\text{ext}} < \frac{1}{3}$,
the sphere becomes a point mass with infinite density at the center. At this stage, the exterior Schwarzschild solution \eqref{Eq 500} becomes relevant. 

It follows from equation \eqref{Eq 602} that for an external observer, a sphere of gravitational mass cannot have a radius measured from outside whose numerical value is smaller than:

\begin{equation} \label{Eq 611}
R_{\text{ext}}= \alpha,    
\end{equation}

\noindent or less than $R =2Gm$ (recall that $\alpha = 2Gm$). If it does have a radius $R_{\text{ext}}< \alpha$, or $R < 2Gm$, then the equations fail, indicating a sphere collapse \cite{Einstein10}. $R = 2Gm$ limits the size of the incompressible fluid sphere to the size $R > 2Gm$. 

\section{Matching the exterior and interior solutions}

Schwarzschild demonstrated that the two solutions \eqref{Eq 514} and \eqref{Eq 500} coincide at the sphere's boundary. He related the radius and density of the sphere to the spherical variables, both inside and at the sphere's surface. Schwarzschild showed that the physical parameters are continuous across the boundary. 
He demonstrated that his specific line element \eqref{Eq 500} accurately described the external space of an incompressible fluid sphere, the vacuum space outside the sphere \cite{Eisenstaedt}. 
This description hinged on the radius vector of the sphere ($r$), which was linked to the sphere's density ($\rho$), and the constant $\alpha$ was associated with the uniform density of the fluid sphere. Accordingly, the external cube of the radius $R_{\text{ext}}^3$ equated to the internal cube of the radius $r_{\text{int}}^3 + \rho$, and also $\alpha$ equaled $\rho$ for a point mass, indicate that the exterior properties of the sphere depend on its density. 

The internal Schwarzschild solution \eqref{Eq 514} should match smoothly with the external Schwarzschild solution \eqref{Eq 500} at the boundary of the sphere. I adhere to the subsequent steps to demonstrate the alignment of the two line elements. 

From the Pythagorean identity, we know that \(\cos^2 \sigma + \sin^2 \sigma = 1\). Hence, equation \eqref{Eq 603} becomes:

\begin{equation} \label{Eq 603a}
v^2_{\text{ext}} = \frac{1}{\cos^2 \sigma_{\text{ext}}}.    
\end{equation}

For an object falling radially (i.e., moving directly towards the center of the sphere without any angular motion), the relevant part of the metric is the radial component. 
Considering only the radial component of the motion and given equation \eqref{Eq 603a} for $\sigma = \sigma_{\text{ext}}$, the line element \eqref{Eq 514} becomes: 

\begin{equation} \label{Eq 604}
ds^2 = \left(\frac{3 \cos \sigma_{\text{ext}} - \cos \sigma}{2}\right)^2 dt^2 - \frac{1}{\cos^2 \sigma}d \sigma^2 + \frac{3}{\kappa \rho_0}{{\sin^2}\sigma} (d\theta^2 + \sin^2\theta d\phi^2).
\end{equation}

Consider the Wikipedia coordinate transformation:\footnote{See \href{https://en.wikipedia.org/wiki/Interior_Schwarzschild_metric}{"Interior Schwarzschild metric"} for more information.} 

\begin{equation} \label{Eq 593}
\sigma = \arcsin\left(\frac{r}{\mathcal{R}}\right),  \text{ and }     
\end{equation}

\begin{equation} \label{Eq 593a}
\sigma_{\text{ext}} = \arcsin\left(\frac{\alpha}{2\mathcal{R}}\right),     
\end{equation}

\noindent which simplifies the line element \eqref{Eq 604}.

\noindent $\mathcal{R}=\frac{r_g^3}{\alpha} \equiv \frac{r_g^3}{r_s}$, where $r_s$ is the Schwarzschild radius. The gravitational radius $r_g$ is given by equation \eqref{Eq 609} (from section \ref{4}):  

\begin{equation} \label{Eq 610}
r_g = \frac{\alpha}{2} \equiv \frac{r_s}{2}.    
\end{equation}

Transformation \eqref{Eq 593} and \eqref{Eq 593a} helps express the interior metric in a form that can easily match the exterior Schwarzschild metric at the boundary. 
Let us do the derivation. 
We will apply the Wikipedia coordinate transformation \eqref{Eq 593} and \eqref{Eq 593a} to the Schwarzschild line element \eqref{Eq 604}, and replace $\sigma$ and $\sigma_{\text{ext}}$ with their respective expressions in terms of $r$, $\mathcal{R}$, and Schwarzschild's notation $\alpha$ or the modern $r_s$. 

Firstly, let us express \(\cos \sigma\) and \(\cos \sigma_{\text{ext}}\) in terms of \(r\) and \(\alpha\). Using the identity \(\cos(\arcsin(x)) = \sqrt{1-x^2}\), we have:

\begin{equation} \label{Eq 600}
\cos \sigma = \cos\left(\arcsin\left(\frac{r}{\mathcal{R}}\right)\right) = \sqrt{1 - \left(\frac{r}{\mathcal{R}}\right)^2},    
\end{equation}

\begin{equation} \label{Eq 600a}
\cos \sigma_{\text{ext}} = \cos\left(\arcsin\left(\frac{\alpha}{2\mathcal{R}}\right)\right) = \sqrt{1 - \left(\frac{\alpha}{2\mathcal{R}}\right)^2},  \text{ and:}   
\end{equation}

\begin{equation} \label{Eq 600b}
\cos^2 \sigma = 1 - \left(\frac{r}{\mathcal{R}}\right)^2.    
\end{equation}

\noindent To transform the expression \(\cos^2 \sigma\) using the transformation equation \eqref{Eq 593}, we need to express \(\cos \sigma\) in terms of \(r\) and \(\mathcal{R}\).
Again using the Pythagorean identity, \(\cos^2 \sigma + \sin^2 \sigma = 1\), since \(\sin \sigma = \frac{r}{\mathcal{R}}\) from the transformation \eqref{Eq 593}, we write:

\[\cos^2 \sigma = 1 - \sin^2 \sigma = 1 - \left(\frac{r}{\mathcal{R}}\right)^2.\]

\noindent This is equation \eqref{Eq 600b}.

\noindent Substituting equations \eqref{Eq 600}, \eqref{Eq 600a}, and \eqref{Eq 600b} back into the line element equation \eqref{Eq 604}, and considering equation \eqref{Eq 596}, we get:

\begin{equation} \label{Eq 594}
ds^2 = \frac{1}{4}\left(3\sqrt{1 - \left(\frac{\alpha}{2\mathcal{R}}\right)^2} - \sqrt{1 - \left(\frac{r}{\mathcal{R}}\right)^2}\right)^2 dt^2  
\end{equation}

\begin{equation*}
-\frac{1}{1-\left(\frac{r}{\mathcal{R}}\right)^2} dr^2 + r^2 \left(d\theta^2 + \sin^2 \theta d\phi^2\right), \qquad \text{if } 0 \leq r \leq r_g.        
\end{equation*}

This equation expresses the metric in terms of the coordinates \(r\), \(\theta\), and \(\phi\), with the parameters \(\mathcal{R}\) and \(\alpha\) representing a specific form of the Schwarzschild interior solution, particularly focused on the boundary condition where $\sigma ={\sigma_{\text{ext}}}$. Equation \eqref{Eq 594} is the resulting line element after applying the coordinate transformation \eqref{Eq 593} and \eqref{Eq 593a}. 

Substituting equation \eqref{Eq 610} into equation \eqref{Eq 594}, we get:

\begin{equation} \label{Eq 594a}
ds^2 = \frac{1}{4}\left(3\sqrt{1 - \left(\frac{r_g}{\mathcal{R}}\right)^2} - \sqrt{1 - \left(\frac{r}{\mathcal{R}}\right)^2}\right)^2 dt^2  
\end{equation}

\begin{equation*}
-\frac{1}{1-\left(\frac{r}{\mathcal{R}}\right)^2} dr^2 + r^2 \left(d\theta^2 + \sin^2 \theta d\phi^2\right), \qquad \text{if } 0 \leq r \leq r_g.    
\end{equation*} 
\vspace{1mm} %1mm vertical space 

At the boundary, the interior solution (describing the spacetime inside the sphere) \eqref{Eq 594a} must match smoothly with the exterior Schwarzschild solution (describing the spacetime outside the sphere):

\begin{equation} \label{Eq 594b}
ds^2 = \left(1 - \frac{r_s}{r}\right) dt^2 -\frac{1}{1-\frac{r_s}{r}} dr^2 + r^2 \left(d\theta^2 + \sin^2 \theta d\phi^2\right),  \text{ if } r_g \leq r.
\end{equation}

\section{Einstein's preference for his approximate solution over Schwarzschild's} \label{9}

In March 1916, Einstein submitted a comprehensive review article on his general theory of relativity, titled "The Foundation of the General Theory of Relativity," to the \emph{Annalen der Physik}, which was then published in May of the same year \cite{Einstein5}. This review paper was written following Schwarzschild's discovery of the exact exterior and interior solutions to Einstein's field equations\eqref{Eq 30} and \eqref{Eq 20}. 
However, in his 1916 paper, Einstein opted for the approximate approach rather than Schwarzschild's exact exterior solution. He chose to discuss the first-order approximate solution he had developed on November 18, 1915, \eqref{Eq 29}, which applies when equation \eqref{Eq 529} is valid:

\begin{equation} \label{Eq 553}
g_{\mu\nu} = 
\begin{cases} 
-(\delta_{\rho\sigma} + \alpha\frac{ x_{\rho} x_{\sigma}}{r^3}) & \text{for } \mu = \rho, \nu = \sigma \text{ and } \rho, \sigma = 1, 2, 3. \\
0 & \text{for } \mu = \rho, \nu = 4 \text{ or } \mu = 4, \nu = \rho \text{ and } \rho = 1, 2, 3. \\
1 - \frac{\alpha}{r} & \text{for } \mu = \nu = 4.
\end{cases}
\end{equation}

\noindent The spatial components $(\rho, \sigma = 1, 2, 3)$ are given by the first case. The mixed space-time components (involving the fourth dimension and any of the first three spatial dimensions) are zero, as stated in the second case. The last case gives the time-time component (see discussion in section \ref{1}). 

In the conclusion of his 1916 review paper on general relativity, Einstein points out that when the gravitational field is calculated with a higher degree of accuracy [beyond the first approximation equation \eqref{Eq 553}], there is a predicted deviation from the Newtonian laws of planetary motion. The specific deviation Einstein refers to is the precession of the elliptical orbit of Mercury. Einstein provides a formula for this precession \cite{Einstein5}:

\begin{equation} \label{Eq 539}
\varepsilon = 24\pi^3 \frac{a^2}{T^2 c^2 (1 - e^2)}.
\end{equation}

\noindent Here, $\varepsilon$ represents the angle of precession or the precession per orbit, $a$ is the semi-major axis of the ellipse, $T$ is the orbital period, $c$ is the speed of light, and $e$ is the eccentricity of the ellipse.

Einstein applies this formula to Mercury and finds that it predicts a precession of the orbit of about $43$ arcseconds per century. He mentions that this result corresponds to observations: astronomers, including Le Verrier, had observed an unexplained precession in Mercury's orbit of approximately the same magnitude. The classical Newtonian mechanics could not account for this anomaly \cite{Einstein5}. 

In a footnote, Einstein refers readers to his original paper on the perihelion of Mercury published in 1915 \cite{Einstein15} and to Schwarzschild's first paper from 1916 \cite{Schwarzschild} for more detailed calculations \cite{Einstein5}. These references indicate where interested readers can find the more technical and detailed aspects of the calculations and solutions he and Schwarzschild developed. However, Einstein does not explicitly utilize Schwarzschild's exact exterior solution equation \eqref{Eq 500} in this derivation \cite{Einstein5}. By referring readers to Schwarzschild's paper \cite{Schwarzschild}, Einstein effectively endorsed the significance of Schwarzschild's exterior solution. This reference played a crucial role in the rapid acceptance and further exploration of Schwarzschild's findings in the scientific community.\footnote{This led to a flurry of activity among mathematicians and physicists who sought to rederive, verify, and extend Schwarzschild's work. These efforts not only underscored the robustness and universality of Schwarzschild's solutions but also paved the way for new insights and developments in general relativity. While this paper focuses primarily on the interplay between Einstein's theories and Schwarzschild's contributions, it is important to acknowledge the broader scientific milieu that actively engaged with and expanded upon these foundational ideas. This subsequent work by various scholars played a crucial role in advancing our understanding of spacetime, black holes, and the very nature of gravity.}

Einstein had already presented equation \eqref{Eq 539} in his 1915 paper on the perihelion of Mercury \cite{Einstein15}.
In his 1915 Mercury perihelion paper, Einstein transforms the following equation:

\begin{equation} \label{Eq 81}
\varepsilon = \pi\frac{3\alpha}{a(1-e^2)},
\end{equation}

\noindent $a$ is the semi-major axis of Mercury's orbit, and $\alpha$ is a constant determined by the sun's mass, into equation \eqref{Eq 539} \cite{Einstein15}. This transformation is an example of how Einstein's general relativity modifies classical mechanics to account for the effects of gravity on space-time. 

Einstein does not provide the exact algebraic steps requiring some work. However, let us transform equation \eqref{Eq 81} into equation \eqref{Eq 539}. Einstein hints about introducing $T$ (orbital period in seconds) \cite{Einstein15}. Indeed, Einstein's relationship involves the square of the period $T^2$, as seen in equation \eqref{Eq 539}. 
We first need to express $\alpha$ in terms of $a, T$ and $c$. This involves Kepler's third law for planetary orbits around the Sun:

\begin{equation} \label{Eq 540}
T^2 = \frac{4\pi^2 a^3}{G M_{\text{sun}}},
\end{equation}

\noindent where $T$ is the orbital period, $G$ is the gravitational constant, and $M_{\text{Sun}}$ is the mass of the Sun.
Recall that $\alpha$ is given by equation \eqref{Eq 73} (see section \ref{4}): $\frac{2GM_{\text{sun}}}{c^2}$.

\noindent Let us use these relations to connect the equations \eqref{Eq 81} and \eqref{Eq 539}. We substitute the expression \eqref{Eq 73} for $\alpha$ into equation \eqref{Eq 81}:

\begin{equation} \label{Eq 542}
\varepsilon = \pi\frac{3\frac{2G M_{\text{sun}}}{c^2}}{a(1-e^2)} = \pi\frac{6G M_{\text{sun}}}{ac^2(1-e^2)}.
\end{equation}

\noindent Next, we use Kepler's third law equation \eqref{Eq 540} to replace $M_{\text{sun}}$. Therefore:

\begin{equation} \label{Eq 541}
M_{\text{sun}} = \frac{4\pi^2 a^3}{G T^2}.
\end{equation}

\noindent We substitute equation \eqref{Eq 541} into the equation for \(\varepsilon\) \eqref{Eq 542}:

\begin{equation}
\varepsilon = \pi\frac{6G \frac{4\pi^2 a^3}{G T^2}}{ac^2(1-e^2)} = \pi\frac{24\pi^2 a^2}{T^2 c^2 (1-e^2)}.
\end{equation}

\noindent Finally, we simplify the equation and obtain Einstein's equation \eqref{Eq 539}.

While Einstein's thinking was fixated on the requirement for unimodular coordinates, he wrote the spatial components of equation \eqref{Eq 553}: 

\begin{equation} \label{Eq 554}
g_{\rho \sigma} = -(\delta_{\rho\sigma} + \alpha\frac{ x_{\rho} x_{\sigma}}{r^3}),    
\end{equation}

\noindent in the following form \cite{Einstein5}:

\begin{equation} \label{Eq 555}
\boxed{g_{11} = -\left(1 + \frac{\alpha}{r}\right).}   
\end{equation}

\noindent $g_{11}$ refers to a spatial component of the metric tensor associated with the radial coordinate in a spherical coordinate system. $g_{11}$ represents the component of the metric tensor associated with the radial dimension (i.e., the $r$ coordinate) in spherical coordinates. It describes how distances are measured along the radial direction. 

\noindent Recall that equation \eqref{Eq 29} incorporates elements typically associated with Cartesian components (as indicated by $x_\rho$ and $x_\sigma$ terms) and spherical coordinates (due to the term $\frac{\alpha}{r}$) (see section \ref{1}) and equation \eqref{Eq 553}. 

Let us derive equation \eqref{Eq 555} from equation \eqref{Eq 554}. To find $g_{11}$, we would need to set $\rho = \sigma = 1$ in equation \eqref{Eq 554}. Assuming that the coordinates are in a standard spherical system (where $1$ represents the radial coordinate),  $x_{\rho} = x_{\sigma} = r$, equation \eqref{Eq 554} becomes:

\begin{equation} \label{Eq 556}
-(\delta_{11} + \alpha\frac{ r^2}{r^3}).    
\end{equation}    

\noindent Since $\delta_{11}=1$, (because $\rho = \sigma$), the equation \eqref{Eq 556} simplifies to:

\begin{equation} 
-(1 + \alpha\frac{1}{r}),     
\end{equation}

\noindent which is Einstein's equation \eqref{Eq 555}.

Recall that the third component $g_{44}$ in equation \eqref{Eq 553} is \cite{Einstein5}:

\begin{equation} \label{Eq 559}
\boxed{g_{44} = \left(1 - \frac{\alpha}{r}\right).}   
\end{equation}

\noindent $g_{44}$, in the third equation of \eqref{Eq 553}, represents a temporal component associated with time in a spacetime metric.

Einstein's first-order approximation solution, as shown in  \eqref{Eq 553}, can be expressed in the following manner \cite{Earman}:

\begin{equation} \label{Eq 511}
ds^2 =   (1 - \frac{\alpha}{r})c^2dt^2  -  \sum_{\rho \sigma} (\delta_{\rho\sigma} + \alpha\frac{ x_{\rho} x_{\sigma}}{r^3}) dx_\rho dx_\sigma. 
\end{equation}

In his perihelion of Mercury paper \cite{Einstein15}, Einstein did not express his approximate solution as a line element. The line element \eqref{Eq 511} describes a spacetime with a correction term to the flat spacetime metric involving the parameter $\alpha$ and coordinates $x_\rho$ and $x_\sigma$. 
The line element \eqref{Eq 511} can be expressed in spherical coordinates \cite{Earman}:

\begin{equation} \label{Eq 512}
ds^2 = (1 - \frac{\alpha}{r}) dt^2 - (1 + \frac{\alpha}{r}) dr^2 - r^2 (d\theta^2 + \sin^2\theta d\phi^2),
\end{equation}

\noindent and compare equation \eqref{Eq 512} to the first-order approximation to the Schwarzschild metric:

\begin{equation} \label{Eq 513}
ds^2 = (1 - \frac{\alpha}{r}) dt^2 - (1 - \frac{\alpha}{r})^{-1} dr^2 - r^2 (d\theta^2 + \sin^2\theta d\phi^2),
\end{equation}

\noindent The term:

\begin{equation*}
\sum_{\rho \sigma} (\delta_{\rho\sigma} + \alpha\frac{ x_{\rho} x_{\sigma}}{r^3}) dx_\rho dx_\sigma  \quad \text{ or: } (1 + \frac{\alpha}{r})dr^2,  
\end{equation*}

\noindent deviates from the first-order approximation to the spatial part of the Schwarzschild metric. The key difference is in the $dr^2$ term.

\noindent Equation \eqref{Eq 559} corresponds to the first term of Equation \eqref{Eq 512}, while Equation \eqref{Eq 555} represents its second term.

As mentioned above, Einstein did not write equation \eqref{Eq 512} in his perihelion of Mercury paper because he imposed the determinant condition \eqref{Eq 529}.
Indeed, the metric tensor $g_{\rho \sigma}$, equation \eqref{Eq 512}: 

\begin{equation} \label{Eq 533}
g_{\rho\sigma} = \begin{pmatrix}
1 + \frac{\alpha}{r} & 0 & 0 & 0 \\
0 & -1 & 0 & 0 \\
0 & 0 & -1 & 0 \\
0 & 0 & 0 & 1 - \frac{\alpha}{r}
\end{pmatrix}
\end{equation}

\noindent does not satisfy the unimodular condition \eqref{Eq 529}.  
The determinant of a diagonal matrix is the product of equation \eqref{Eq 512}'s diagonal elements. So, the determinant $g$ of matrix \eqref{Eq 533} is:

\begin{equation}
g = \left(1+\frac{\alpha}{r}\right) \times (-1) \times (-1) \times \left(1-\frac{\alpha}{r}\right)= \frac{\alpha^2}{r^2} - 1.
\end{equation}

\noindent \noindent We will calculate this determinant and its square root to check if the unimodular condition is met.
The square root of the negative determinant of the metric tensor is:

\begin{equation}
\sqrt{-g} = \sqrt{1 - \frac{\alpha^2}{r^2}}.
\end{equation}

\noindent This expression does not simplify to $1$. Therefore, the metric tensor \eqref{Eq 512} does not satisfy the unimodular condition \eqref{Eq 529}.
\vspace{1mm} %1mm vertical space  

$g_{11}$ in the equation \eqref{Eq 555} influences the measurement of distances in the radial direction. It modifies how lengths and distances are calculated near a massive object, and using equation \eqref{Eq 555}, Einstein calculates length contraction \cite{Einstein5}. On the other hand, $g_{44}$ equation \eqref{Eq 559} affects the measurement of time intervals. It encapsulates how time dilates in a gravitational field. 

Let us reverse-engineer Einstein's calculation using equation \eqref{Eq 555} to show how the presence of a gravitational field alters the measurement of distances, leading to the conclusion of length contraction. Consider the line element:

\begin{equation}
ds^2 = g_{\mu \nu} dx_\mu dx_\nu.    
\end{equation}

\noindent When a unit-measuring rod is laid parallel to the $x$-axis, only the $dx_1$ component (along the $x$-axis) is non-zero; the other components $dx_2, dx_3, dx_4$ are zero. Einstein writes the condition $ds^2 = -1$ (a space-like interval), which implies the rod has a proper length of $1$ unit in this spacetime geometry. Therefore, the line element simplifies to \cite{Einstein5}:

\begin{equation} \label{Eq 558}
-1 = g_{11}dx^2_1.    
\end{equation}

\noindent It represents the spatial distance measured along the $x$-axis. Substituting $g_{11}$, the metric component equation \eqref{Eq 555}, into the simplified line element equation \eqref{Eq 558}, we get: 

\begin{equation}
-1 = -(1 + \frac{\alpha}{r})dx^2_1.    
\end{equation}

\noindent We remove the negative signs on both sides of the equation and rearrange the equation to isolate $dx^2_1$ on one side:

\begin{equation}
dx^2_1 =  \frac{1}{1 + \frac{\alpha}{r}}.  
\end{equation}

\noindent To solve for $dx_1$, we take the square root of both sides. Therefore, we get:

\begin{equation}
dx_1 =  \sqrt{\frac{1}{1 + \frac{\alpha}{r}}}.    
\end{equation}

\noindent The expression inside the square root can be rewritten as:

\begin{equation}
dx_1 =  \frac{1}{\sqrt{{1 + \frac{\alpha}{r}}}}.    
\end{equation}

\noindent Since we are considering a unit measuring rod and assuming small $\frac{\alpha}{r}$,  we can expand this to the first order. For a small $\frac{\alpha}{r}$ the square root function $\sqrt{{1 + \frac{\alpha}{r}}}$ can be expanded using a Taylor series around $\frac{\alpha}{r}= 0$. 

\noindent The result is the following equation \cite{Einstein5}:

\begin{equation} \label{Eq 557}
dx_1 \approx  1 - \frac{1}{2}\frac{\alpha}{r}.    
\end{equation} 

\noindent Equation \eqref{Eq 557} shows that in a weak gravitational field (where $\frac{\alpha}{r}$ is small), the effect of the field on the length measured in the direction of the field ($x$ direction) is a slight contraction, as expected from the principles of general relativity.

\section{Why did Einstein stick to the approximation approach?} \label{10}

Schwarzschild's solution \eqref{Eq 500}, derived shortly after Einstein's theory was published in 1915, offered the first exact solution to Einstein's field equations \eqref{Eq 30} and \eqref{Eq 20}, providing a precise and comprehensive description of the spacetime geometry of the mass point according to Einstein's four conditions. However, when addressing Mercury's perihelion advance in his 1916 review paper,  "The Foundation of the General Theory of Relativity," Einstein employed an approximate solution equation \eqref{Eq 553} within the framework of unimodular coordinates. Why did Einstein persist in using his approximate methods in the form \eqref{Eq 554} in his 1916 paper \cite{Einstein5}? Einstein's preference for using his approximate methodology can be understood in the context of the physical and mathematical framework he employed in his 1916 review article: 

It appears very likely that Einstein's focus on approximate solutions had little to do with his later reservations about the singularity inherent in Schwarzschild's solution. In 1916, the full implications of such singularities were not well understood, and the concept of a black hole as we know it today had not yet been developed. In his early correspondence with Schwarzschild, Einstein did not address the singularities inherent in Schwarzschild's solution \cite{CPAE8}, Doc. 176, Doc. 181, Doc. 194. This lack of mention likely indicates that the significance of the singularities in the Schwarzschild line element was not fully realized at the time or that Einstein's immediate concerns were elsewhere, particularly in validating and further developing his 1915 theory of general relativity. It is well-known that Einstein had reservations about singularities, as suggested by the Schwarzschild metric. However, as said above, these reservations became more pronounced later in his career as the physical and mathematical implications of the Schwarzschild metric became clearer.

Einstein persisted in using his approximate methods in the form \eqref{Eq 553} in his 1916 paper \cite{Einstein5}. This persistence can be attributed partly to the inherent complexity of exact solutions and the practical difficulties in applying them to various problems within the theory. Additionally, Einstein's strong inclination towards unimodular coordinates significantly influenced this choice. Unimodular coordinates, fundamental to his earlier work, especially in calculating Mercury’s perihelion advance, seemed to offer a more manageable and intuitive framework for his theoretical explorations. Despite the advancements represented by Schwarzschild's exact solutions, this fixation on unimodular coordinates underscores the complexity and evolving nature of Einstein's theory development, reflecting a balance between theoretical elegance and practical applicability. In March 1916, when Einstein submitted his review paper on general relativity, the condition for unimodular coordinates \eqref{Eq 529} was something that he found challenging to move beyond and detach himself from because unimodular coordinates offered simplifications in the mathematical formulation of his theory. In 1915, it appeared to him that working within the unimodular coordinates allowed for a clearer physical interpretation of general covariance. The Schwarzschild metric in spherical coordinates \eqref{Eq 500} does not satisfy the condition for unimodular coordinates \eqref{Eq 529}.

Angular momentum or the area law, equation \eqref{Eq 508}, can be deduced from the line element in equation \eqref{Eq 512}  -- a formulation that Einstein himself did not provide -- through the application of Euler-Lagrange equations \cite{Earman}. On the other hand, the derivation of conservation laws like equation \eqref{Eq 508} using Einstein's approximate methodology is less straightforward. While Einstein's conditions 1) to 4) (time independence, zero mixed components, spherical symmetry, and solution vanishes at infinity, see section \ref{2}) include spherical symmetry, expressing this symmetry in Cartesian coordinates is less direct than in spherical coordinates. Schwarzschild showed that the clear and direct derivation of angular momentum conservation (or the area law) relies on the explicit spherical symmetry present in spherical coordinates, where the lack of dependency on the azimuthal angle $\phi$ leads straightforwardly to the conservation of angular momentum. Applying the Euler-Lagrange equations to derive conservation laws requires additional assumptions, especially considering the Cartesian framework. The conservation laws are not as immediately apparent as in the Schwarzschild solution, where the spherical symmetry and specific metric form directly lead to such laws.

Unlike Einstein's methodology, Schwarzschild approached the problem more mathematically. His solution to Einstein's field equations was exact and did not rely on the heuristic methods that Einstein used.
Schwarzschild's solution provided a rigorous mathematical description of the spacetime around a spherical mass. Schwarzschild's exact solution possibly challenged the framework Einstein had used. 
In 1916, Einstein's theory of general relativity was still new and developing. The full implications of Schwarzschild's solution to Einstein's field equations were not immediately understood and appreciated. The potential to derive the deflection of light by gravity from the Schwarzschild metric was not immediately recognized. The connection between light deflection and the Schwarzschild solution was not obvious in 1916. The broader implications of Schwarzschild's solution were not grasped in 1916. Over time, as the understanding of general relativity deepened, the significance of Schwarzschild's solution became clearer. It was pivotal in developing the theory of (non-rotating) black holes and studying strong gravitational fields. And Einstein grew uncomfortable with the idea of singularities, as suggested by Schwarzschild's solution.

Einstein's approach to general relativity was characterized by a series of iterative refinements and pragmatic approximations deeply rooted in physical principles. His early attempts to describe the gravitational field around a static sun were more heuristic and less exact than Schwarzschild's precise solution. 
His use of approximate methods, especially in unimodular coordinates, was instrumental in deriving key predictions of general relativity, such as the perihelion advance of Mercury, the deflection of light by gravity, and the contraction of lengths. 
Einstein relied on fundamental principles to guide his theoretical development rather than solely on rigorous mathematical derivation. His approximations with the condition for unimodular coordinates were especially relevant in scenarios where the gravitational fields were weak, and the deviations from the Newtonian limit were small.
Einstein's estimation of the approximate form of the gravitational field around the Sun represented a Newtonian interpretation, suggesting the Sun's gravitational field was not much different from the one considered by Laplace. However, this approach sufficiently addressed the issue of Mercury's perihelion precession \cite{Eisenstaedt}.

In February 1916, Schwarzschild suggested to Einstein finding a condition different from the determinant condition equation \eqref{Eq 527} that could further simplify the field equations of general relativity \cite{CPAE8}, Doc 188. 
Einstein's response to Schwarzschild's letter about two and a half weeks later contained the following explanations \cite{CPAE8}, Doc. 194: in the Addendum to his first paper on general relativity, the November 4, 1915 paper \cite{Einstein6}, published on November 11, 1915 \cite{Einstein6}, Einstein revised the restriction to unimodular transformations to the condition for unimodular coordinates equation \eqref{Eq 529}. Einstein justified the change by stating that his comments in the November 4, 1915 paper \cite{Einstein1} were outdated. 
In this first paper on general relativity \cite{Einstein1}, Einstein wrestled with the weak field approximation. He postulated that $\sqrt{-g}=1$, so this condition was a starting point for his theory. It was a foundational assumption on which the rest of the theoretical framework was built. He imposed the condition related to the harmonic condition \cite{Einstein1}:

\begin{equation} \label{227}
\sum_\beta\left(\frac{\partial g^{\alpha\beta}}{\partial x_\alpha}\right) = 0,    
\end{equation}

\noindent and derived the equation:

\begin{equation} \label{Eq 228} 
\frac{1}{2} \sum_{\alpha} \frac{\partial^2 g_{\mu\nu}}{\partial x_{\alpha}^2} = \kappa T_{\mu\nu},    
\end{equation}

\noindent from his November 4. 1915 field equations. These equations retained their form under unimodular transformations.  

In his letter addressed to Schwarzschild, Einstein elucidated the following points. The strategy of selecting a coordinate system based on the coordinate condition equation \eqref{227} \cite{Einstein1} is incompatible with postulating and setting $\sqrt{-g}=1$. 
Following this insight, Einstein told Schwarzschild that his approach to integrating Newtonian principles within the comprehensive framework of the theory has evolved. After the November 4, 1915 paper (see \cite{Einstein6}), in his subsequent papers \cite{Einstein15}, \cite{Einstein14} and review paper on general relativity \cite{Einstein5}, Einstein shifted to an alternative approach when dealing with the Newtonian limit, consistent with his field equations \eqref{Eq 20} valid only in unimodular coordinates. 

After imposing the unimodular coordinate condition \eqref{Eq 529}, Einstein focused on the Newtonian limit using an approach more suitable for drawing parallels with Newtonian mechanics, especially for weak gravitational fields and the Newtonian approximation.
In contrast, the other condition \eqref{227} aligns more with a harmonic gauge, which is useful in wave solutions and special relativistic analogies.

Eventually, Einstein recognized that a departure from unimodular coordinates was necessary to explore the concept of gravitational waves. The harmonic condition (related to the equation \eqref{227}) was more appropriate for analyzing wave-like solutions in the theory, as it facilitated the comparison with Maxwell's equations of electromagnetism.
However, Einstein's journey in developing and refining his theory was marked by revisions, reconsiderations, and occasional missteps. After initially exploring and setting aside the unimodular coordinate condition in his work on gravitational waves, Einstein revisited this condition. This revisit was partly due to an error he made in his calculations. Einstein's involvement and subsequent complications with the unimodular coordinate condition culminated in him informing Schwarzschild that this could imply the non-existence of gravitational waves, similar to light waves \cite{CPAE8}, Doc. 194. See discussion in \cite{Kennefick}.

It is well known that Einstein aimed for simplicity and broader accessibility in his general relativity. He was searching for a system that was as simple and logical as possible and with minimum axioms. Using an approximation methodology to explain the precession of Mercury's orbit and the bending of light provided a clear, understandable, and compelling argument for the validity of his 1916 general relativity. 
When Einstein developed his theory between 1912 and 1916, the mathematical tools and understanding of differential geometry were still evolving. As he developed his theory, Einstein learned these concepts from Marcel Grossmann and other mathematicians like Tullio Levi-Civita. Furthermore, in his quest to formulate the correct form of the field equations, Einstein sought to ensure that his field equations were consistent with the conservation of energy and momentum and the correspondence principle (the Newtonian limit).  
The factors of mathematical tools and the two principles probably influenced Einstein's preference for unimodular transformations and the subsequent condition for unimodular coordinates, which he felt comfortable manipulating. Despite his initial preference for unimodular coordinates, Einstein recognized the importance of generalizing his theory beyond any specific coordinate system, even in his 1916 review paper. A note and passage in the paper \cite{Einstein5} and manuscript indicate his awareness of the need to move beyond any particular coordinate choice, including unimodular coordinates. Indeed, this awareness marked a pivotal point in Einstein's theoretical journey, leading him to move beyond unimodular coordinates in pursuit of a more generalized framework (see discussion in my paper \cite{Weinstein2}). 

\section*{Acknowledgement}

This work is supported by ERC advanced grant number 834735.

\end{document}